\documentclass[trackchanges, twocolumn]{aastex701}

 \usepackage{xcolor}

\usepackage{amsmath}

\usepackage{placeins}
\received{March 20, 2026}
\revised{May 25, 2026}
\accepted{May 28, 2026}
\submitjournal{The Astrophysical Journal}
\reportnum{arXiv:2603.09876}

\begin{document}

\title{The rotational and magnetic properties of Polaris from long-term spectropolarimetric monitoring}

\author[orcid=0000-0003-2088-0706,sname='Barron']{James A. Barron}
\affiliation{Department of Physics, Engineering Physics \& Astronomy, Queen’s University, 64 Bader Lane, Kingston, ON K7L 3N6, Canada}
\email[show]{j.barron@queensu.ca}  

\author[orcid=0000-0002-1854-0131,sname='Wade']{Gregg A. Wade} 
\affiliation{Department of Physics \& Space Science, Royal Military College of Canada, PO Box 17000, Station Forces, Kingston, ON K7K 7B4, Canada}
\email{NA}

\author[orcid=0000-0002-9023-7890,sname=Folsom]{Colin P. Folsom}
\affiliation{Tartu Observatory, University of Tartu, Observatooriumi 1, Tõravere, 61602, Estonia}
\email{NA}

\begin{abstract}
Polaris is a highly unusual Cepheid with observed properties that are difficult to reconcile with stellar evolutionary models. Since the initial detection of Polaris' magnetic field in 2020, we have conducted a magnetic monitoring campaign with the ESPaDOnS spectropolarimeter at the Canada-France-Hawaii Telescope. We compute Stokes~$V$ least-squares deconvolution profiles and measure the associated mean longitudinal magnetic field strengths $\langle B_{z}\rangle$. The surface magnetic field has remained remarkably stable over five years of observations, with $\langle B_{z}\rangle$ varying between approximately $-3$\,G and $+0.6$\,G. From the periodic modulation of $\langle B_{z}\rangle$ we infer a stellar rotation period of $P_{\mathrm{rot}}=100.29\pm0.19$\,days. This is the first direct measurement of $P_{\mathrm{rot}}$ for a classical Cepheid. Previous interferometric radius measurements and $P_{\mathrm{rot}}$ imply an equatorial rotation velocity of $v_{\mathrm{eq}}=23.3\pm0.2$\,km\,s$^{-1}$. We set a conservative upper bound on the projected equatorial rotational velocity of $v_{\mathrm{eq}}\sin i_{\star} < 13.5$\,km\,s$^{-1}$ and constrain the stellar inclination angle to be $i_{\star}<37^{\circ}$. Using the previously determined orbital solution, we find a high likelihood of a strong spin-orbit misalignment. We determine the lower bound on the obliquity angle between the stellar rotation and orbital axes to be $\beta>18.7^{\circ}$ at 99\% confidence. We discuss the challenges in interpreting the origin and properties of the surface magnetic field in the context of Polaris' uncertain evolutionary history and the merger hypothesis. 
\end{abstract}



\section{Introduction} \label{sec:Introduction}
As the pole star, \object{Polaris} ($\alpha$ UMi, \object{HD 8890}) is one of the best known stars in the night sky. It is the nearest and brightest classical Cepheid (hereafter Cepheid), and has been the subject of extensive study since the mid 1800's (e.g. \citealt{Seidel_1852, Campbell_1899, Hertzsprung_1911, Roemer_1965, Ferro_1983, Turner_2005, Anderson_2019, Torres_2023}). Keen interest in Polaris has been driven by puzzling changes in its variability and challenges in reconciling its inferred physical properties with stellar pulsation and evolution models.

\cite{Barron_2022} reported the first high-resolution, circularly polarized spectroscopic observation of Polaris as part of first results from a large magnetic survey of Cepheids. A Stokes~$V$ Zeeman magnetic signature was detected at high signal-to-noise ratio (S/N), with a corresponding weak mean longitudinal magnetic field measured with sub-gauss precision. It is becoming clear that Cepheid surface magnetic fields are frequently detectable in (ultra)-high S/N spectropolarimetric observations (\citealt{Grunhut_2010, Barron_2022, Barron_2024}, in prep.). However, we know little about the origins of these fields and their relation to pulsation dynamics.    

We have conducted a long-term spectropolarimetric monitoring campaign of Polaris using the Echelle SpectroPolarimetric Device for the Observation of Stars (ESPaDOnS) at the Canada-France-Hawaii-Telescope (CFHT). Polaris is an ideal Cepheid target for magnetic monitoring due to its bright apparent magnitude, which permits high-precision magnetic measurements to be acquired in reasonable integration times. Additionally, Polaris' Stokes~$V$ profiles show less distortion and appear easier to interpret compared to Cepheids such as $\delta$~Cep or $\eta$~Aql \citep{Barron_2022}. 

In stars with non-axisymmetric magnetic fields, the Zeeman induced Stokes~$V$ signatures and corresponding $\langle B_{z}\rangle$ measurements are periodically modulated by the stellar rotation as different regions of the star's magnetic geography are seen by the observer \citep{Donati_2009}. This makes stellar magnetometry an effective method to directly measure rotation periods, particularly in slowly rotating stars that do not present unambiguous rotationally modulated signals in spectral or photometric time-series (e.g. Cepheids). 

Rotation is a fundamental stellar property, impacting evolutionary tracks and surface chemical abundances \citep{Meynet_2000, Ekstrom_2012}. Cepheid blue loops are sensitive to the initial rotation rate at the zero-age main-sequence (ZAMS). This suggests that rotation is a key ingredient in resolving the long-standing Cepheid mass discrepancy problem \citep{Anderson_2014, Anderson_2016, Zhao_2023}, alongside other processes including convective core overshooting \citep{Cassisi_2011, Prada_Moroni_2012}, and pulsation driven mass loss \citep{Neilson_2011}.

Furthermore, rotation and magnetic field evolution are interconnected processes. Rotation is essential for the generation of both convective and radiative dynamos (\citealt{Kapyla_2023} and references therein). In turn, coupling between the magnetic field and stellar wind leads to angular momentum loss and stellar spindown \citep{Weber_1967, ud_Doula_2009}.

Recent results necessitate a direct measurement of Polaris' rotation period. An apparent age discrepancy between the Cepheid Polaris Aa and its distant companion Polaris B has led to the suggestion that the Cepheid may be a merger product \citep{Bond_2018, Evans_2018, Anderson_2018}. This is an intriguing possibility, considering the growing evidence that mergers are a viable channel to produce strong magnetic fields in intermediate and massive stars \citep{Schneider_2016, Schneider_2019, Korvakova_2022, Shenar_2023, Frost_2024, Korvakova_2025, Schneider_2025}. Recent binary population synthesis and N-body modelling predict that a significant fraction of Cepheids are the product of binary interactions, including coalescence \citep{Dinnbier_2024}. In addition to Polaris, there is a growing number of Cepheid merger candidates identified in binary systems \citep{Neilson_2015, Pilecki_2022, Pilecki_2024, Espinoza_2025}. A constraint on Polaris' rotation rate at the ZAMS in concert with recent interferometric radius and dynamical mass measurements \citep{Evans_2024} will permit detailed comparison to evolution models and testing of the merger hypothesis.  

\cite{Evans_2024} reported the detection of surface spots on Polaris from an interferometric surface image reconstruction using the CHARA Array. The nature of these surface features and their possible relation to the magnetic field geometry is unclear. Relatedly, there is ambiguity surrounding the origin and stability of Polaris' long spectral periods ($\gtrsim 40$\,days; \citealt{Bruntt_2008, Anderson_2019, Barbey_2025}), which may be due to rotating surface features \citep{Dinshaw_1989, Lee_2008} or non-radial pulsations \citep{Hatzes_2000}.

In this paper, we present results from five years of spectropolarimetric monitoring of Polaris. In Sec.~\ref{sec:Polaris} we summarize Polaris' physical and variability properties, in Sec.~\ref{sec:Observations} we discuss the ESPaDOnS observations and in Sec.~\ref{sec:Magnetic_Analysis} we present the magnetic analysis. In Sec.~\ref{sec:Variability_Analysis} we perform a variability analysis on the RV and $\langle B_{z}\rangle$ measurements and determine the stellar rotation period. In Sec.~\ref{sec:rot_and_inc} we determine broadening parameters from spectral fitting and constrain the stellar inclination angle. In Sec.~\ref{sec:spin_orbit} we constrain the spin-orbit misalignment of the inner binary Polaris~Aa. In Sec.~\ref{sec:Discussion} we discuss our results, and in Sec.~\ref{sec:Conclusions} we list our conclusions.

\section{Polaris}\label{sec:Polaris}
Polaris Aa (F8\,Ib, $\langle V\rangle=2.0$\,mag) the Cepheid is a member of a confirmed triple system. The inner single-lined spectroscopic binary consists of Polaris Aa in a $\sim30$\,year eccentric orbit (\citealt{Torres_2023} and references therein) with Polaris~Ab (F6\,V; \citealt{Evans_2008}). Separated at $\sim18\arcsec$, Polaris B is a wide, gravitationally bound visual companion of spectral type $\sim\mathrm{F3}$\,V \citep{Turner_1977, Evans_2008, Usenko_2008}.

Polaris has a pulsation period of $P_{\mathrm{puls}}\approx3.97$\,days, and shows remarkably low-amplitude, sinusoidal variations in radial-velocity (RV) measurements and photometry. More than a century of spectroscopic and photometric monitoring has revealed seemingly irregular changes in the pulsation period and amplitude over long time scales (e.g. \citealt{Torres_2023}). Proposed origins for this behaviour include magneto-convective cycles \citep{Stothers_2009}, pulsation instabilities \citep{Evans_2002} or tidal perturbations induced by the binary companion at periastron \citep{Torres_2023}.

Polaris’ pulsation mode and instability strip (IS) crossing number have been the subjects of vigorous debate \citep{Neilson_2012, Turner_2013, Neilson_2014, Fadeyev_2015, Bond_2018, Anderson_2018, Evans_2018, Evans_2024}. The conflicting interpretations are largely due to disagreement about Polaris' distance before \textit{Gaia} \citep{Engle_2018}. \cite{Evans_2024} adopt a \textit{Gaia} DR3 distance of $136.90\pm0.34$\,pc to the Polaris system, anchored to Polaris B. This value is close to the \textit{Hipparcos} distance of $132.6\pm1.9$\,pc for Polaris A \citep{van_Leeuwen_2007, van_leeuwen_2013}. At this distance Polaris is considered a first overtone Cepheid from comparison to the period-luminosity and period-radius relations for fundamental Cepheids \citep{Feast_1997, Neilson_2014, Evans_2018}. A first overtone pulsation mode is also supported by Fourier analysis of the RV curve, which is independent of the distance measurement \citep{Hocde_2024}.

The most recent spectroscopic studies of Polaris report an effective temperature $T_{\mathrm{eff}}\approx6000$\,K and surface gravity $\log g\approx2.0$ \citep{Usenko_2005, Ripepi_2021}. \cite{Evans_2024} report a dynamical mass measurement of $M_{\star}=5.13\pm0.28\,M_{\odot}$ and a mean interferometric radius of $\langle R_{\star}\rangle=46.27\pm0.42\,R_{\odot}$.

Polaris A is a soft X-ray source with a luminosity ($\log L_{X}\approx28.9$\,erg\,s$^{-1}$) and spectrum that could be produced by either Polaris Aa or Ab \citep{Evans_2010, Engle_2015, Evans_2022}. Polaris~A's X-ray flux does not show phase dependant enhancement seen in the higher amplitude Cepheids $\delta$~Cep and $\beta$~Dor \citep{Engle_2015, Evans_2022}. A circumstellar envelope (CSE) of unclear origin has been detected at $\sim2.4\,R_{\star}$, comprising $\sim1.5$\% of Polaris' $K$-band flux \citep{Merand_2006}. As the temperature at this distance is too hot for dust formation, the CSE is may be due to free-free emission from an ionized gas shell \citep{Hocde_2020}. 

\section{Observations} \label{sec:Observations}
We obtained spectropolarimetric observations of Polaris with ESPaDOnS at CFHT \citep{Donati_2006}. The observations were acquired between November 2020 and December 2025 under program IDs 20BC20, 21BC39, 22AC07, 22BC17, 23AC11, 23BC25, 24BC27 and 25BD03 (PI: Barron). ESPaDOnS is a bench-mounted, cross-dispersed echelle spectrograph with a spectral range of $370-1050$\,nm and a resolving power of $R\sim65,000$. Each spectropolarimetric sequence consists of four sub-exposures between which the Fresnel rhombs are rotated in the Cassegrain-mounted polarimeter. The sequences were reduced using the Upena pipeline (based on Libre-ESpRIT; \citealt{Donati_1997}) to produce one unpolarized Stokes~$I$ spectrum, one circularly polarized Stokes $V$ spectrum and two diagnostic null spectra. 

Multiple short-exposure sequences were obtained and co-added as Polaris quickly saturates the charge-coupled device due to its bright apparent magnitude. We excluded poor quality polarimetric sequences that were affected by poor observing conditions or technical issues, leaving 42 co-added observations for analysis. The spectra were combined and rectified using \texttt{normPlot} from the SpecpolFlow software package \citep{Folsom_2025}. The total integration time of the co-added spectra was reduced from $\sim2500$\,s for the 2020 observation to $\sim500$\,s for subsequent observations. The median S/N of all 42 co-added spectra is $\sim3000$ at 500\,nm (per 1.8\,km\,s$^{-1}$ pixel). Table~\ref{tab:obs} provides a log of all co-added observations.

\section{Magnetic Analysis}\label{sec:Magnetic_Analysis} 
We performed a magnetic analysis similar to that described by \cite{Barron_2022}. We used the SpecpolFlow software package \citep{Folsom_2025} to generate least-squares-deconvolution (LSD) profiles \citep{Donati_1997, Kochukhov_2010}. The LSD profiles were computed with a velocity binning of 1.8\,km\,s$^{-1}$ using the tailored line mask generated by \cite{Barron_2022} from 400\,nm to 900\,nm. The adopted LSD normalization parameters are $\lambda_{0}=500$\,nm, $g_{0}=1.2$ and $d_{0}=0.1$ for the wavelength, effective Land\'e factor and line depth respectively. All LSD Stokes~$V$ profiles are definitely detected using the false alarm probability (FAP) metric and thresholds described by \cite{Donati_1992, Donati_1997}. No spurious signatures are detected in the LSD $N$ profiles. \cite{Barron_2022} previously ruled out instrumental cross-talk from linear polarization as a contributor to Polaris' Stokes~$V$ signatures.

We measured the centre-of-gravity radial velocity ($\mathrm{RV}_{\mathrm{cog}}$) of each LSD Stokes~$I$ profile using the first-moment method as implemented in SpecpolFlow. Each LSD profile was shifted by the corresponding $\mathrm{RV}_{\mathrm{cog}}$ to correct for pulsation and orbital motion. The mean longitudinal magnetic field $\langle B_{z}\rangle$ for each observation was calculated from the first moment of the LSD~Stokes~$V$ profile \citep{Donati_1997, Wade_2000} using
\begin{equation}
    \langle B_{z}\rangle = -2.14\times10^{11}\frac{\int vV(v)dv}{\lambda_{0}g_{0}c\int[1-I(v)dv]}
\end{equation}
where $c$ is the speed of light and $I(v)$ and $V(v)$ are the continuum normalized Stokes~$I$ and $V$ profiles. The $\langle B_{z}\rangle$ calculations were performed using constant integration bounds between $-30$\,km\,s$^{-1}$ and $+30$\,km\,s$^{-1}$ on the $\mathrm{RV}_{\mathrm{cog}}$ corrected LSD profiles with SpecpolFlow. We find that $\langle B_{z}\rangle$ varies between approximately $-3$\,G and $+0.6$\,G with a typical uncertainties of $\sim0.3$\,G. The $\mathrm{RV}_{\mathrm{cog}}$ and $\langle B_{z}\rangle$ measurements are reported in Table~\ref{tab:obs}.

We re-measured $\langle B_{z}\rangle=0.28\pm0.13$\,G for the 2020 observation  compared to $\langle B_{z}\rangle=0.59\pm0.16$\,G reported by \cite{Barron_2022}. This difference is primarily due to our adoption of a slightly smaller integration range and the exclusion of a single poor quality spectropolarimetric sequence from the co-added observation.
\begin{figure}
    \centering
    \includegraphics[width=\linewidth]{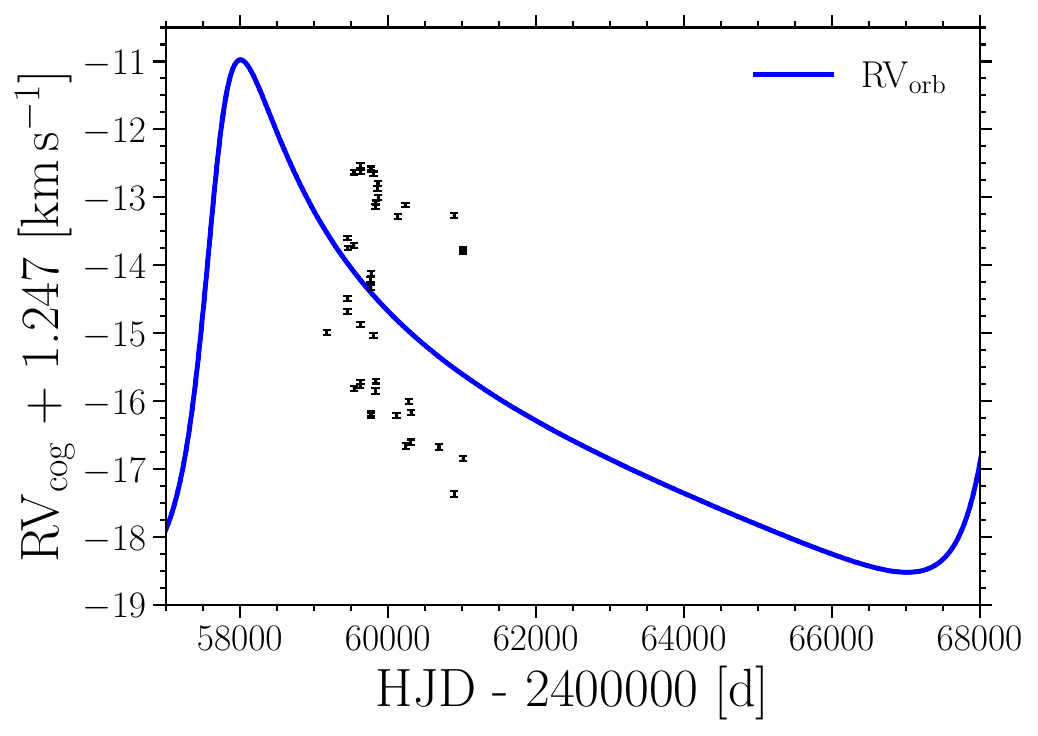}
    \caption{Offset corrected $\mathrm{RV}_{\mathrm{cog}}$ measurements (black markers) shown in relation to the best fit orbital RV curve (solid blue line) given by \cite{Evans_2024}. The scatter in $\mathrm{RV}_{\mathrm{cog}}$ is due to pulsation.}
    \label{fig:orbit}
\end{figure}

\section{Variability Analysis}\label{sec:Variability_Analysis}
\subsection{Pulsation Period and Amplitude}\label{sec:Pulsation_Period_Amplitude}
Figure~\ref{fig:orbit} shows the $\mathrm{RV}_{\mathrm{cog}}$ measurements in relation to orbital RV curve ($\mathrm{RV}_{\mathrm{orb}}$). We subtracted the systemic velocity and orbital motion from the $\mathrm{RV}_{\mathrm{cog}}$ measurements using the ``replicated" weighted orbital solution determined by \cite{Evans_2024}. After subtraction an offset of approximately $-1.25$\,km\,s$^{-1}$ remained as \cite{Evans_2024} calibrated the zero-points of different RV data sets against the data presented by \cite{Roemer_1965}. 

We modelled the orbit-subtracted $\mathrm{RV}_{\mathrm{cog}}$ measurements with a first-order sine function of the form 
\begin{equation}\label{eq:rv_model}
    \mathrm{RV}_{\mathrm{cog}} - \mathrm{RV}_{\mathrm{orb}} = C + A\sin\left(\frac{2\pi}{P_{\mathrm{puls}}}(t-t_{0})\right)
\end{equation}
where $C$ is the systematic offset from the reference frame of the orbital solution and $A$ is the semi-amplitude. We chose the reference epoch $t_{0}$ to coincide with the time of minimum radius nearest the middle of the observation span. This sets the pulsation phase $\phi_{\mathrm{puls}}=0$ to correspond to a pulsation-averaged velocity of 0\,km\,s$^{-1}$ on the descending branch of the pulsation curve. We determined best fit values and uncertainties for parameters in Eq.~\ref{eq:rv_model} using a Markov Chain Monte Carlo (MCMC) analysis described in Appendix~\ref{sec:MCMC}. We also experimented with fitting a two harmonic Fourier model similar to \cite{Anderson_2024}. However, we found that the inclusion of additional terms did not significantly alter the best fit model.

\begin{figure}
    \centering
    \includegraphics[width=\linewidth]{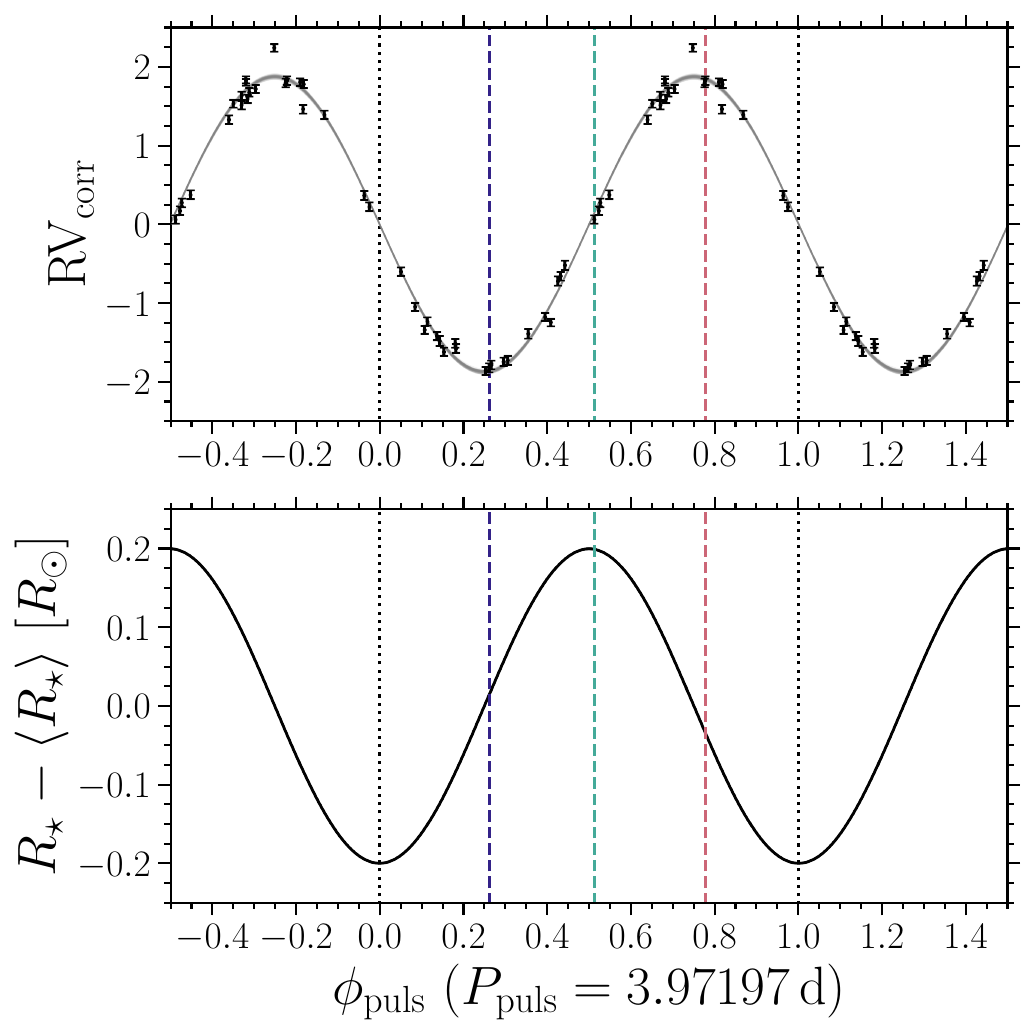}
    \caption{\textbf{Top}: Orbit and offset corrected RV measurements shown over two pulsation cycles. The grey lines show 100 random samples drawn from the MCMC joint posterior distribution. \textbf{Bottom}: Variation in radius inferred from the best fit pulsation model. The  black dotted lines correspond to $R_{\mathrm{min}}$ and the dashed coloured lines correspond to the LSD profiles shown in Fig.~\ref{fig:LSD_StokesIV_pulsation}.}
    \label{fig:RVcorr_radius}
\end{figure}

We determined $P_{\mathrm{puls}}=3.97197(4)$\,days, which is consistent with $P_{\mathrm{puls}}=3.972013(13)$\,days reported by \cite{Anderson_2024}. The pulsation ephemeris over our observation span is 
\begin{equation}\label{eq:puls_ephem}
    \mathrm{HJD} = 2460095.714^{+0.005}_{-0.005}+ 3.97197^{+0.00004}_{-0.00004}\cdot E
\end{equation}
We determined an unsigned semi-amplitude of $|A|=1.87\pm0.01$\,km\,s$^{-1}$ and offset $C=-1.247\pm0.008$\,km\,s$^{-1}$. We define the orbit- and offset-corrected RV measurements to be $\mathrm{RV}_{\mathrm{corr}}=\mathrm{RV}_{\mathrm{cog}}-\mathrm{RV}_{\mathrm{orb}}+1.247$. Figure~\ref{fig:RVcorr_radius} shows $\mathrm{RV}_{\mathrm{corr}}$ and the best fit model as a function of $\mathrm{\phi}_{\mathrm{puls}}$. Values of $\mathrm{RV}_{\mathrm{corr}}$ and $\phi_{\mathrm{puls}}$ are recorded in Table~\ref{tab:obs}.

The RV amplitude is largely consistent with measurements reported since 2020 \citep{Buecke_2021, Torres_2023, Usenko_2024}, and the longer increasing trend observed since 1990 when the peak-to-peak amplitude reached a minimum of $\sim1.5$\,km\,s$^{-1}$. However, the current pulsation amplitude remains low, even for an overtone Cepheid \citep{Anderson_2024}.

The stellar radius as a function of time is calculated from 
\begin{equation}\label{eq:radius}
    R(t) = \langle R_{\star}\rangle - p\int\mathrm{RV_{\mathrm{corr}}(t)}\,dt
\end{equation}
where $p$ is the projection factor \citep{Storm_2011}. Polaris' full amplitude of radius variation ${\Delta R=R_{\mathrm{max}}-R_{\mathrm{min}}}$ was determined by \cite{Moskalik_2005} to be $\Delta R\approx0.17$\,$R_{\odot}$. As the pulsation amplitude has increased over the past 20 years, we calculated a revised estimate of $\Delta R\approx 0.4$\,$R_{\odot}$ from integration of the best fit $\mathrm{RV}_{\mathrm{corr}}$ model. We adopted $p=1.36$ to facilitate direct comparison with \cite{Moskalik_2005}. The lower panel of Fig.~\ref{fig:RVcorr_radius} shows the radius variation as a function of $\phi_{\mathrm{puls}}$. 

While Cepheid $p$-factors can range over $\sim1.1-1.5$, precise determinations are challenging as $p$ does not have a clear correlation with $P_{\mathrm{puls}}$ and other stellar parameters \citep{Pilecki_2018, Trahin_2021}. Our primary purpose for calculating $\Delta R$ is to obtain an order of magnitude estimate for the change in magnetic field strength due to the pulsation (Sec.~\ref{sec:mag_field_and_pulsation}). This estimation is not sensitive to our choice of $p$. 

\begin{figure}[h!]
    \centering
    \includegraphics[width=\linewidth]{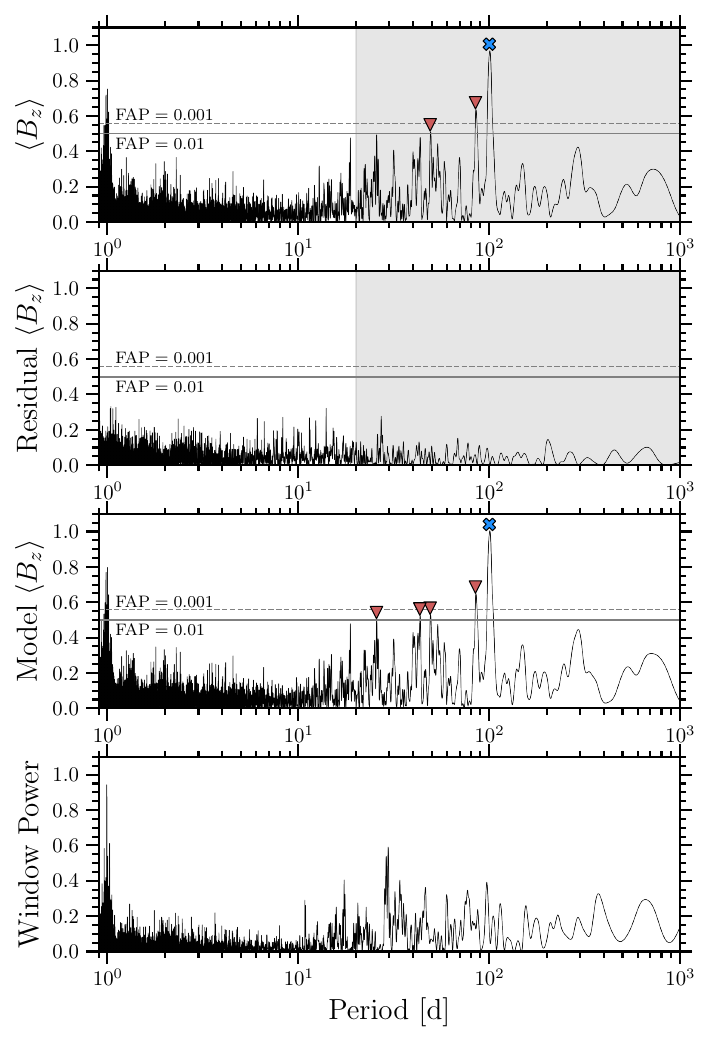}
    \caption{From top to bottom: GLS periodogram of $\langle B_{z}\rangle$ measurements, GLS periodogram of $\langle B_{z}\rangle$ residuals after subtraction of best fitting model, GLS periodogram of model $\langle B_{z}\rangle$ curve and window power spectrum. The solid and dashed grey lines denote the $0.01$ and $0.001$ FAP levels respectively. The blue `X' denotes $P_{\mathrm{rot}}$ and red triangles denote long period aliases. Grey shaded regions indicate allowed rotation periods above $P_{\mathrm{rot}}^{\mathrm{min}}$. }
    \label{fig:GLS_Bz}
\end{figure}

\subsection{Rotation Period}\label{sec:rot_period}
 The critical stellar rotation velocity is given by 
\begin{equation}
v_{\mathrm{crit}}=\sqrt{\frac{2}{3}\frac{GM_{\star}}{R_{\mathrm{p,crit}}}}    
\end{equation}
where $R_{\mathrm{p,crit}}$ is the polar radius at $v_{\mathrm{crit}}$ \citep{Ekstrom_2008}. Adopting $R_{\mathrm{p,crit}}=R_{\mathrm{min}}$ gives $v_{\mathrm{crit}}\approx120$\,km\,s$^{-1}$. The estimate of $v_{\mathrm{crit}}$ sets a conservative minimum stellar rotation period of $P_{\mathrm{rot}}^{\mathrm{min}}\approx20$\,days. 
 
 We searched for significant periods above $P_{\mathrm{rot}}^{\mathrm{min}}$ in the $\langle B_{z}\rangle$ measurements using the generalized Lomb-Scargle (GLS) periodogram \citep{Lomb_1976, Scargle_1982, Zechmeister_2009, vanderplas_2018}. The GLS and window power were computed using the \texttt{LombScargle} class in the \texttt{astropy} Python package \citep{astropy_2022}, and the FAP levels were estimated using the \cite{Baluev_2008} approximation. Figure~\ref{fig:GLS_Bz} shows the GLS periodogram and window power. The highest peak at $\sim100.3$\,days and a peak at ${\sim84.8}$\,days are detected with $\mathrm{FAP}<0.001$. An additional period at ${\sim49.1}$\,days is detected with $\mathrm{FAP}<0.01$. Significant peaks clustered near 1\,day are due to the daily alias. 

\begin{figure*}
    \centering
    \includegraphics[width=\linewidth]{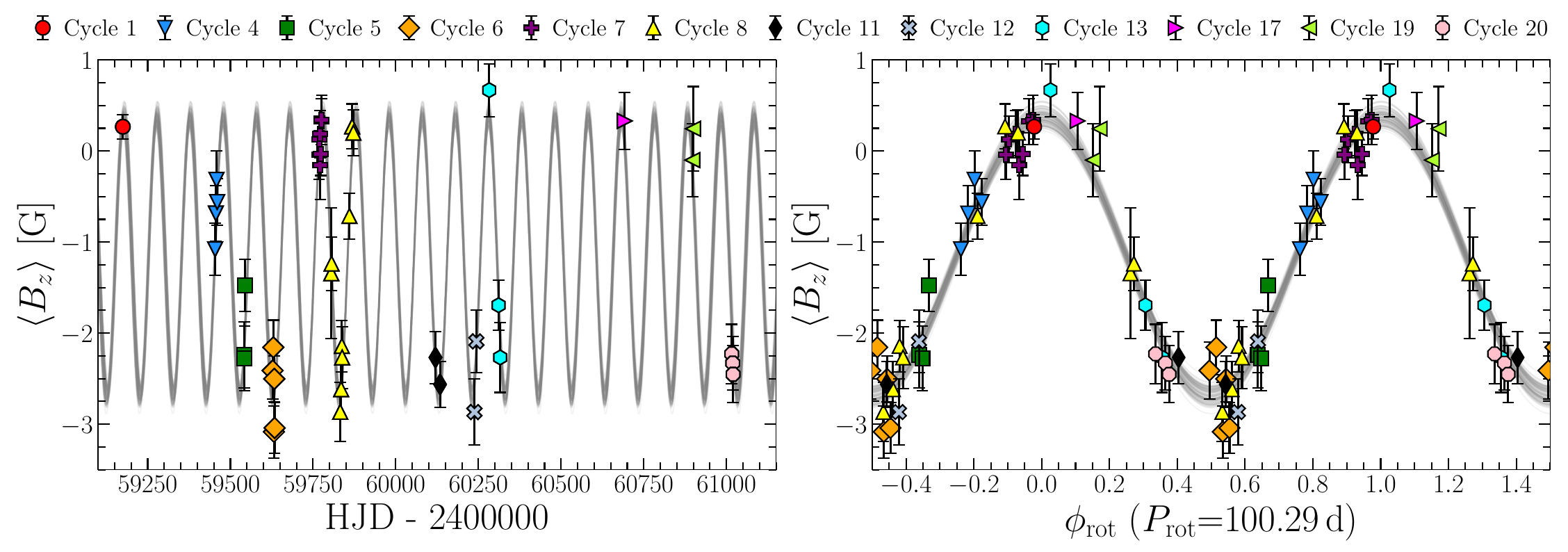}
    \caption{Polaris $\langle B_{z} \rangle$ measurements plotted as a function of HJD (left) and $\phi_{\mathrm{rot}}$ (right). The marker shapes and colours denote $N_{\mathrm{rot}}$. The grey lines show 100 random samples drawn from the MCMC joint posterior distribution.}
    \label{fig:Bz_2plot}
\end{figure*}

The second row of Figure~\ref{fig:GLS_Bz} shows the residual $\langle B_{z}\rangle$ GLS periodogram after subtraction of the best fit model. No significant periods are detected in the residuals, suggesting that $\sim84.8$\,days and $\sim49.1$\,days are aliases of $P_{\mathrm{rot}}$. We confirmed this by computing the GLS periodogram of the $\langle B_{z} \rangle$ model fit with the same time sampling and uncertainties as the observations. The model $\langle B_{z} \rangle$ power spectrum (Fig.~\ref{fig:GLS_Bz}, third row) is nearly identical to the power spectrum of the $\langle B_{z} \rangle$ measurements. Long period aliases are detected at $\sim84.8$\,days, $\sim49.1$\,days, $\sim43.3$\,days and $\sim25.7$\,days ($\mathrm{FAP}<0.01$). 

For a stable magnetic field geometry, LSD~Stokes~$V$ profiles obtained over multiple rotational cycles will show similar morphologies at similar rotation phases. As an additional check, we visually verified that the alias periods yield LSD~Stokes~$V$ profiles with inconsistent morphologies at similar phases. 

The $\langle B_{z}\rangle$ measurements are well modelled by a first-order cosine function of the form 
\begin{equation}\label{eq:cosine}
    \langle B_{z} \rangle = B_{0} + B_{1}\cos\left(\frac{2\pi}{P_{\mathrm{rot}}}(t-t_{0})\right)
\end{equation}
where $B_{0}$ is the mean of the $\langle B_{z}\rangle$ variation and and $B_{1}$ is the semi-amplitude. We chose $t_{0}$ to correspond to the epoch nearest the middle of the observing span where $\langle B_{z} \rangle$ is maximum. Under this formulation the rotation phase $\phi_{\mathrm{rot}}=0$ corresponds to maximum $\langle B_{z} \rangle$.  

\begin{figure*}
    \centering
    \includegraphics[width=\linewidth]{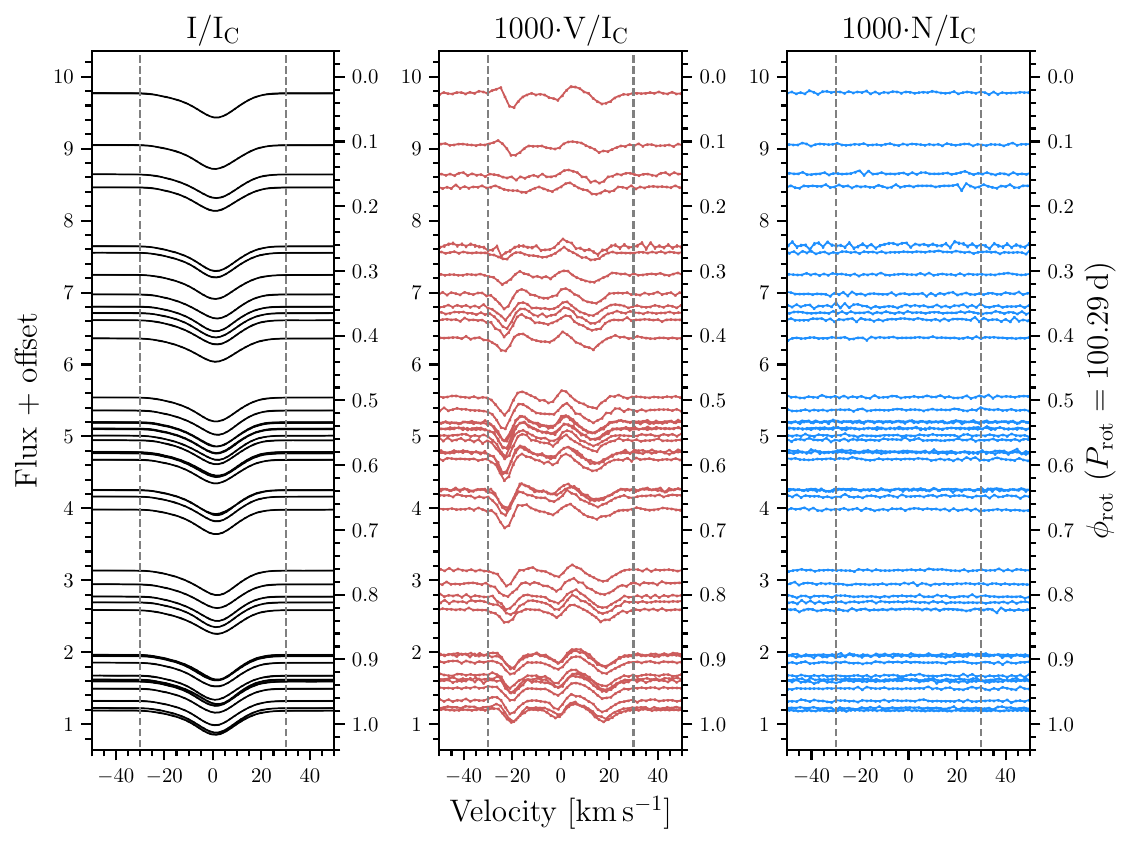}
    \caption{Continuum normalized and $\mathrm{RV}_{\mathrm{cog}}$ corrected LSD profiles of Polaris plotted as a function of $\phi_{\mathrm{rot}}$. From left to right, LSD Stokes~$I$ (black), LSD Stokes~$V$ (red) and LSD $N$ (blue). The vertical dashed gray lines mark the integration bounds in the FAP and $\langle B_{z}\rangle$ analysis.}
    \label{fig:LSD_StokesIVN}
\end{figure*}

To estimate robust parameter uncertainties, we fit Eq.~\ref{eq:cosine} to the $\langle B_{z} \rangle$ measurements using a MCMC analysis described in Appendix~\ref{sec:MCMC}. We obtained ${P_{\mathrm{rot}}=100.29\pm0.19}$\,days, ${B_{0}=-1.15\pm0.06}$\,G and ${B_{1}=1.52\pm0.06}$\,G. The rotation ephemeris given by
\begin{equation}\label{eq:rot_ephem}
    \mathrm{HJD} = 2460079.59^{+0.83}_{-0.83} + 100.29^{+0.19}_{-0.19}\cdot E
\end{equation}
Table~\ref{tab:obs} provides $\phi_{\mathrm{rot}}$ and rotation cycle number $N_{\mathrm{rot}}$ calculated from Eq.~\ref{eq:rot_ephem} for each observation. Figure~\ref{fig:Bz_2plot} shows the $\langle B_{z} \rangle$ measurements as a function of observation date and phased to Eq~\ref{eq:rot_ephem}. Figure~\ref{fig:LSD_StokesIVN} shows the LSD Stokes~$I$, $V$ and $N$ profiles phased to Eq~\ref{eq:rot_ephem}. 

\subsection{Comparison to Literature Periods}\label{sec:alias}
 The $100.3$\,days signal is inconsistent with previously reported spectroscopic periodicities, including ${\sim60/120}$\,days which have been suggested as possible rotation periods \citep{Lee_2008, Anderson_2019, Barbey_2025}. We phased the $\langle B_{z}\rangle$ measurements to the six long periods identified by \cite{Barbey_2025}, as well as $\sim40.2$\,days \citep{Hatzes_2000, Anderson_2019}. None of these literature periods produced coherent $\langle B_{z}\rangle$ curves.

These results strengthen our interpretation of ${\sim100.3}$\,days as the stellar rotation period. We do not observe any other significant periodicity in the $\langle B_{z}\rangle$ data that is not an alias. The $\langle B_{z}\rangle$ variation is incompatible with previously published periods, including $P_{\mathrm{puls}}$, and the variation has remained stable over $\sim5$\,years of observations. 

\subsection{Magnetic Field Strength and Pulsation}\label{sec:mag_field_and_pulsation}
\begin{figure}[h!]
    \centering
    \includegraphics[width=\linewidth]{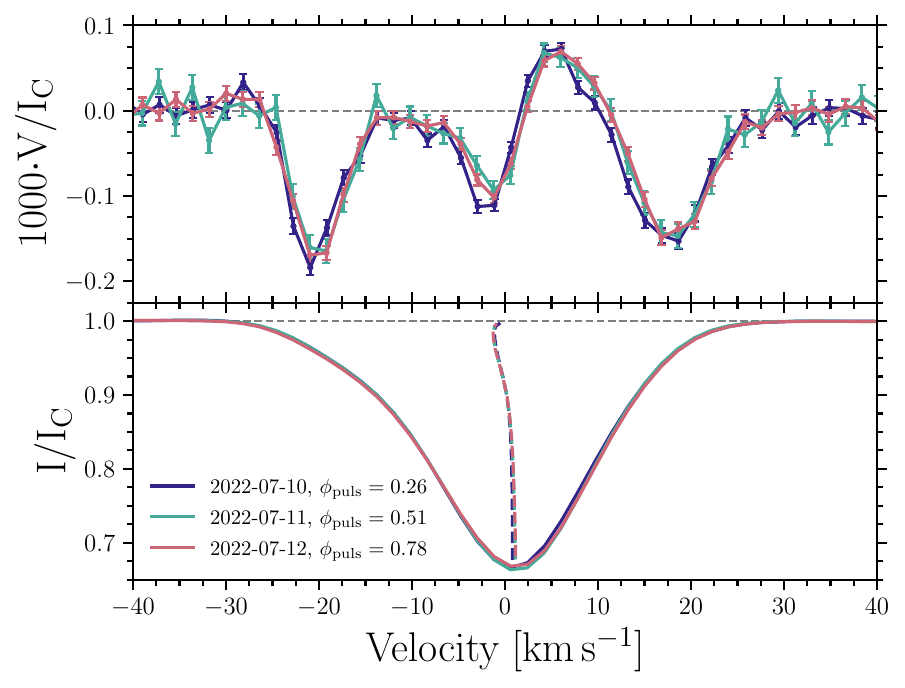}
    \caption{Continuum normalized and $\mathrm{RV}_{\mathrm{cog}}$ corrected LSD~Stokes~$V$ (top) and $I$ (bottom) profiles over three consecutive nights. The Stokes~$I$ bisectors are shown by the vertical dashed lines. The observations sample the radial expansion from minimum to maximum radius (see Fig.~\ref{fig:RVcorr_radius}). No strong variability is observed.}
    \label{fig:LSD_StokesIV_pulsation}
\end{figure}

We do not observe significant changes in the LSD~Stokes~$V$ profiles due to pulsation. Figure~\ref{fig:LSD_StokesIV_pulsation} shows Stokes~$I$ and $V$ profiles obtained over three consecutive nights. The profiles span the ascending branch of the RV curve as the atmosphere expands from average radius to maximum radius and then contracts. No strong variability is observed in the LSD profiles. This is consistent with the absence of significant power near $P_{\mathrm{puls}}\approx3.97$\,days in the $\langle B_{z}\rangle$ GLS periodogram (Fig~\ref{fig:GLS_Bz}). 

The insensitivity of the Stokes~$V$ profiles to pulsation phase is not entirely unexpected given Polaris' low-amplitude pulsation. Under an assumption of magnetic flux conservation, a $\langle B_{z}\rangle$ modulation would not be detectable at the current amplitude. The surface magnetic field strength is related to the radius by
\begin{equation}\label{eq:cons_mag}
    \frac{B_{R_{\mathrm{max}}}}{B_{R_{\mathrm{min}}}} = \left(\frac{R_{\mathrm{min}}}{R_{\mathrm{max}}} \right)^{2}
\end{equation}
where $B_{R_{\mathrm{max}}}$ and $B_{R_{\mathrm{min}}}$ are the surface magnetic field strengths at maximum radius $R_{\mathrm{max}}$ and minimum radius $R_{\mathrm{min}}$ respectively. From the radius variation calculated in Sec.~\ref{sec:Pulsation_Period_Amplitude} we have $R_{\mathrm{min}}\approx46.1$\,$R_{\odot}$ and $R_{\mathrm{max}}\approx46.5$\,$R_{\odot}$. We define the cycle averaged surface field strength be ${\langle B_{\mathrm{puls}}\rangle=\frac{1}{2}(B_{R_{\mathrm{max}}} + B_{R_{\mathrm{min}}})}$. For $\langle B_{\mathrm{puls}}\rangle<10$\,G, Eq.~\ref{eq:cons_mag} gives a total variation of $<0.2$\,G. This variation is too low to be detectable as it is below the precision of the $\langle B_{z}\rangle$ measurements.

\subsection{LSD Stokes~$I$ Bisectors}\label{sec:BIS}
\begin{figure}
    \centering
    \includegraphics[width=\linewidth]{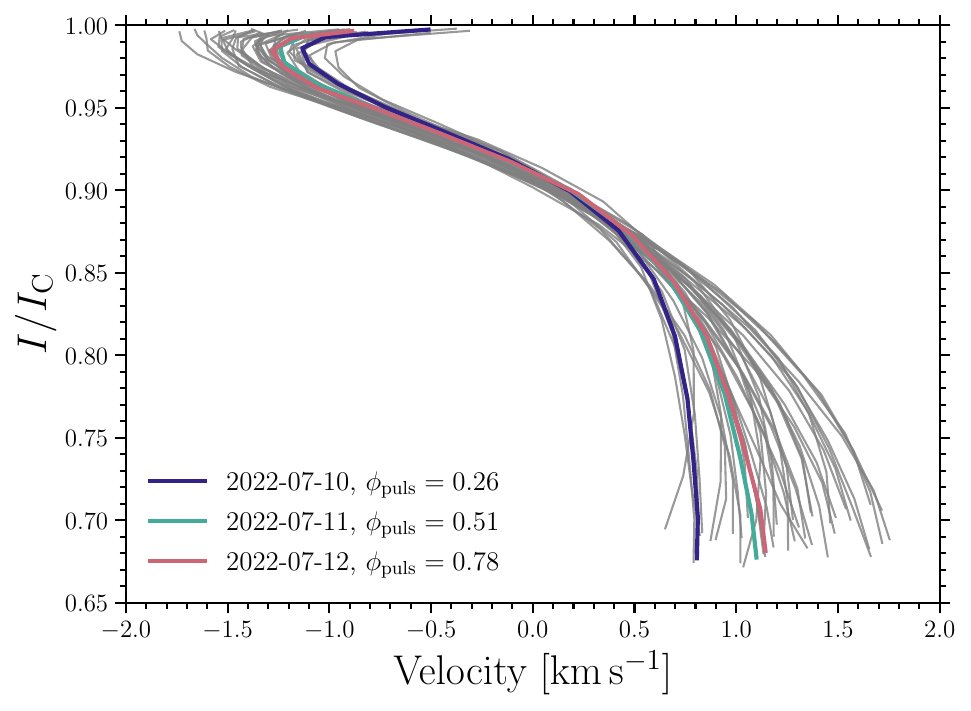}
    \caption{Bisector variability in the $\mathrm{RV}_{\mathrm{cog}}$ corrected LSD~Stokes~$I$ profiles. The coloured lines correspond to the LSD profiles shown in Fig.~\ref{fig:LSD_StokesIV_pulsation}.}
    \label{fig:LSD_bisector}
\end{figure}
We calculated the bisector \citep{Gray_1982} and associated bisector inverse span (BIS; \citealt{Queloz_2001}) for each $\mathrm{RV}_{\mathrm{cog}}$-corrected LSD~Stokes~$I$ profile. The calculations were performed using the \texttt{calc\_bis} routine implemented in SpecpolFlow. The BIS is defined to be $\mathrm{\mathrm{BIS}}=\overline{v}_{t}-\overline{v}_{b}$ where $\overline{v}_{t}$, $\overline{v}_{b}$ are the average velocities calculated respectively from the top $10–40\%$ and bottom $60–90\%$ of the Stokes~$I$ profile. Figure~\ref{fig:LSD_bisector} shows the bisector variability across all observations.

\begin{figure}
    \centering
    \includegraphics[width=\linewidth]{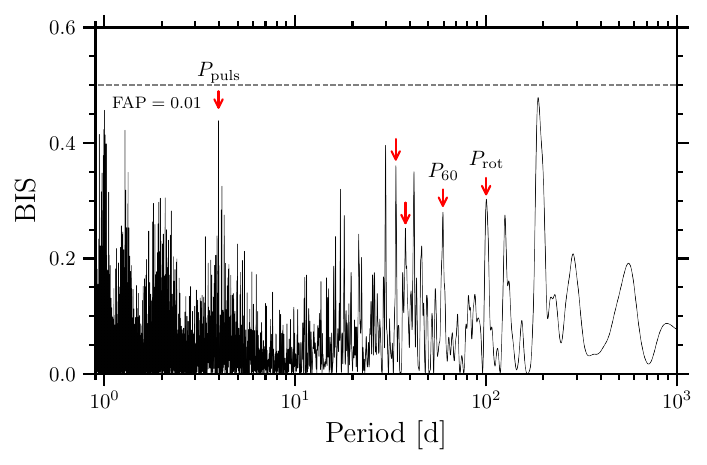}
    \caption{GLS periodogram of the BIS measurements. The horizontal dashed line denotes $\mathrm{FAP}=0.01$. No significant periods are detected. Peaks that correspond to $P_{\mathrm{puls}}$, $P_{\mathrm{rot}}$ and $P_{60}\approx60$\,days are labelled and marked with red arrows. The unlabelled arrows mark $\sim33.7$\,days and $\sim37.9$\,days.}
    \label{fig:GLS_BIS}
\end{figure}

We performed a period analysis on the BIS measurements similar to the $\langle B_{z}\rangle$ GLS analysis described in Sec.~\ref{sec:rot_period}. Figure~\ref{fig:GLS_BIS} shows the BIS GLS periodogram. No significant periods are detected at the $\mathrm{FAP}=0.01$ level. While not formally detected, peaks corresponding to $P_{\mathrm{puls}}$ and $P_{\mathrm{rot}}$ can be clearly identified. The peak at $\sim59.5$\,days appears consistent with reports of a $P_{60}\approx60$\,day period \citep{Anderson_2019, Barbey_2025} and half the 119.1\,days RV period reported by \cite{Lee_2008}. Additionally, peaks at $\sim33.7$\,days and $\sim37.9$\,days are close to candidate signals reported by \cite{Barbey_2025}.   
\section{Rotation, Line Broadening and Stellar Inclination}\label{sec:rot_and_inc}
  With a direct measurement of $P_{\mathrm{rot}}$, the equatorial rotation velocity at mean radius is given by
\begin{equation}
    v_{\mathrm{eq}} = \frac{2\pi \langle R_{\star} \rangle}{P_{\mathrm{rot}}}.
\end{equation}
For $\langle R_{\star} \rangle=46.27\pm0.42\,R_{\odot}$ \citep{Evans_2024} and $P_{\mathrm{rot}}=100.29\pm0.19$\,days we obtain a precise measurement of ${v_{\mathrm{eq}} = 23.3\pm0.2}$\,km\,s$^{-1}$. 

In principle the stellar inclination angle $i_{\star}$ is determinable from a spectroscopic measurement of the projected equatorial rotation velocity $v_{\mathrm{eq}}\sin i_{\star}$. The challenges in disentangling broadening mechanisms in slowly rotating, radial pulsators has been previously explored for Cepheids \citep{Bersier_1996, GIllet_1999, Gray_2007} and RR Lyraes \citep{Peterson_1996, Preston_2019}. Line asymmetries due to pulsation, velocity gradients and granulation complicate the separation of macroturbulent broadening $v_{\mathrm{mac}}$ and low $v_{\mathrm{eq}}\sin i_{\star}$. We performed a spectral broadening analysis to investigate the correlation between $v_{\mathrm{eq}}\sin i_{\star}$ and $v_{\mathrm{mac}}$ and the subsequent impact on our ability to constrain $i_{\star}$.

The broadening analysis was performed on the co-added $\mathrm{RV}_{\mathrm{cog}}$-corrected {2022-07-10} spectrum which was obtained at a $\phi_{\mathrm{puls}}$ near mean radius (Fig.~\ref{fig:RVcorr_radius}). We first applied the line-depth ratio (LDR) method \citep{Kovtyukh_2000} to measure ${T_{\mathrm{eff}}=6030\pm10}$\,K. We excluded LDRs that utilized the Ni\,\textsc{ii}\,$\lambda5755$ and Cr\,\textsc{i}\,$\lambda6330$ lines due to telluric contamination. Our $T_{\mathrm{eff}}$ value is consistent with previous LDR measurements \citep{Usenko_2005, Turner_2013, Ripepi_2021}. The stellar mass and radius \citep{Evans_2024} imply a surface gravity of $\log g=1.82\pm0.03$. 

Model Stokes~$I$ spectra were synthesized using the \texttt{ZEEMAN} spectrum synthesis code \citep{Landstreet_1988, Wade_2001}. We adopted a grid of plane-parallel ATLAS9 atmosphere models \citep{Kurucz_1970, Kurucz_1993, Castelli_2003} assuming local thermodynamic equilibrium. The rotational broadening in \texttt{ZEEMAN} is implemented  by applying local Doppler shifts to the sub-divided stellar disc. The code utilizes a radial-tangential model for $v_{\mathrm{mac}}$ \citep{Gray_1975, Gray_2022}. The spectral fitting was performed using a Levenberg-Marquardt $\chi^{2}$ minimization procedure \citep{Press_1992} described by \cite{Folsom_2012}. 

While our assumption of LTE is not strictly optimal for Cepheids due to their dynamic, low density atmospheres \citep{Bergemann_2025}, it is a sufficient approximation for our analysis. Departures from LTE primarily modify spectral line depths, impacting inferred atmospheric parameters and chemical abundances \citep{Przybilla_2021, Nunnari_2026}. However, as we are primarily interested in the spectral line broadening, the effects of non-LTE on line depths can be safely ignored. Additionally, non-LTE effects are mitigated by Polaris' low pulsation amplitude \citep{Vasilyev_2019}. 

We first performed a simultaneous fitting of three spectral regions: $5500-5673\,\mathrm{\AA}$, $6050-6176\,\mathrm{\AA}$ and $6365-6432\,\mathrm{\AA}$. These regions were chosen to minimize the impacts of severe line blending and telluric contamination. To help break degeneracies between parameters we fixed $T_{\mathrm{eff}}=6030$\,K and $\log g=1.82$. We allowed individual elemental abundances and the RV to vary. The best fit values for the broadening parameters were $v_{\mathrm{eq}}\sin i_{\star}=1.3\pm1.4$\,km\,s$^{-1}$, $v_{\mathrm{mac}}=14.8\pm0.2$\,km\,s$^{-1}$ and microturbulence $v_{\mathrm{mic}}=4.14\pm0.02$\,km\,s$^{-1}$.

\begin{figure}
    \centering
    \includegraphics[width=\linewidth]{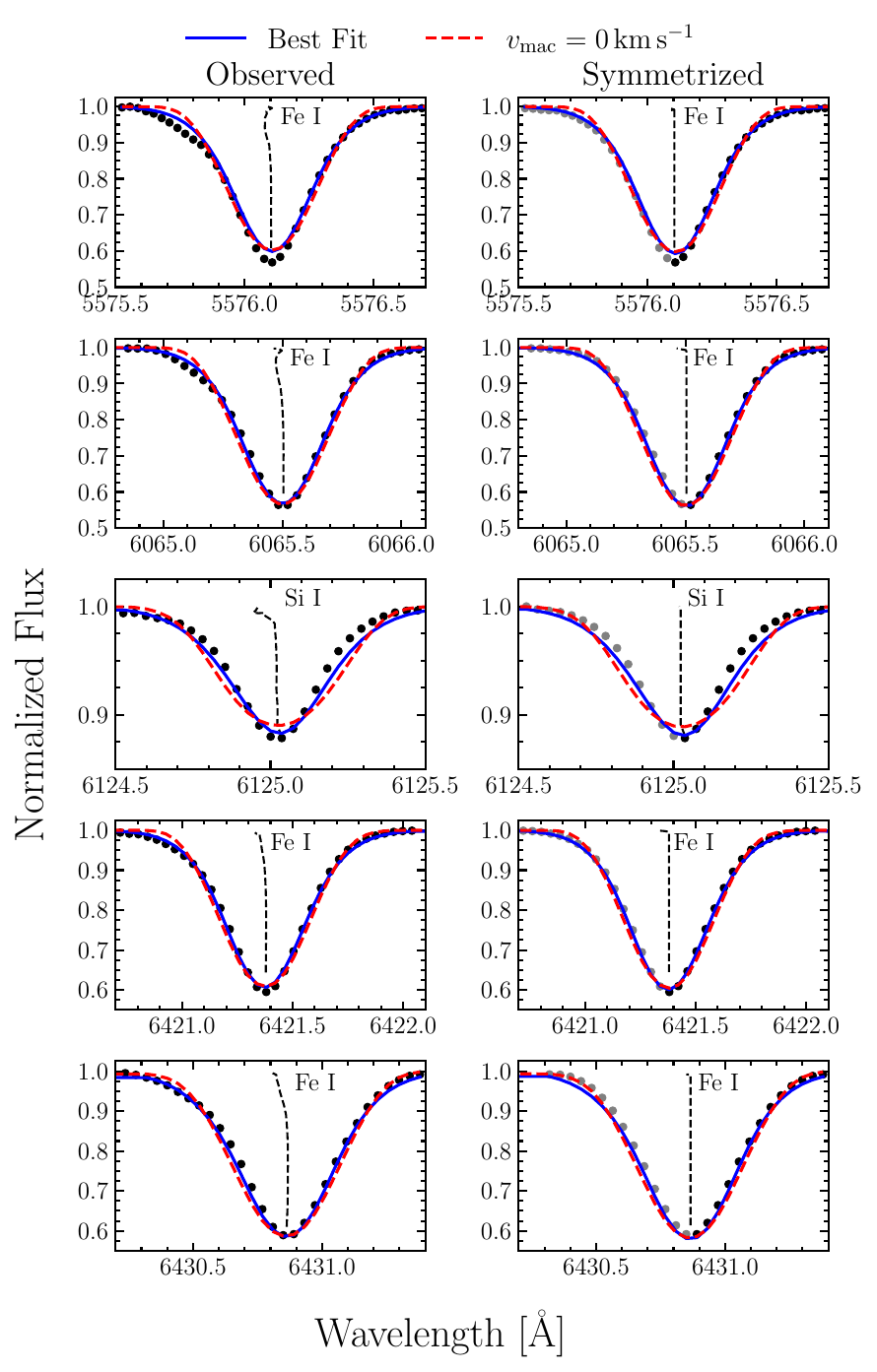}
    \caption{Model fits to the five observed (left column) and symmetrized (right column) line profiles. Grey points denote interpolated flux values in the symmetrized profiles. The bisectors are represented by black dashed lines. The solid blue lines denote the best fit models for $v\sin i$ free and $v_{\mathrm{mac}}$ free. The dashed red lines denote model fits for $v\sin i$ free and $v_{\mathrm{mac}}=0$\,km\,s$^{-1}$.}
    \label{fig:5lines_fit}
\end{figure}

We then selected five metal lines to perform a detailed broadening analysis: Fe\,\textsc{i}\,$\lambda5576$, Fe\,\textsc{i}\,$\lambda6066$, Fe\,\textsc{i}\,$\lambda6125$, Si\,\textsc{i}\,$\lambda6421$ and Fe\,\textsc{i}\,$\lambda6431$. The chosen lines are well reproduced by the best fit model, do not have significant blends or telluric contamination and have depths less than 50\% of the continuum. The line profiles and associated bisectors are shown in the left column of Fig~\ref{fig:5lines_fit}. Similar to the LSD~Stokes~$I$ profiles, the selected lines all show some degree of asymmetry in the blue wing. To verify that our results were not sensitive to the asymmetries, we symmetrized the line profiles by interpolating the red wing onto the wavelength grid of the blue wing. The symmetrized lines are shown in the right column of Fig~\ref{fig:5lines_fit}. 

\begin{figure*}
    \centering
    \begin{minipage}{0.48\textwidth}
        \centering
        \includegraphics[width=\linewidth]{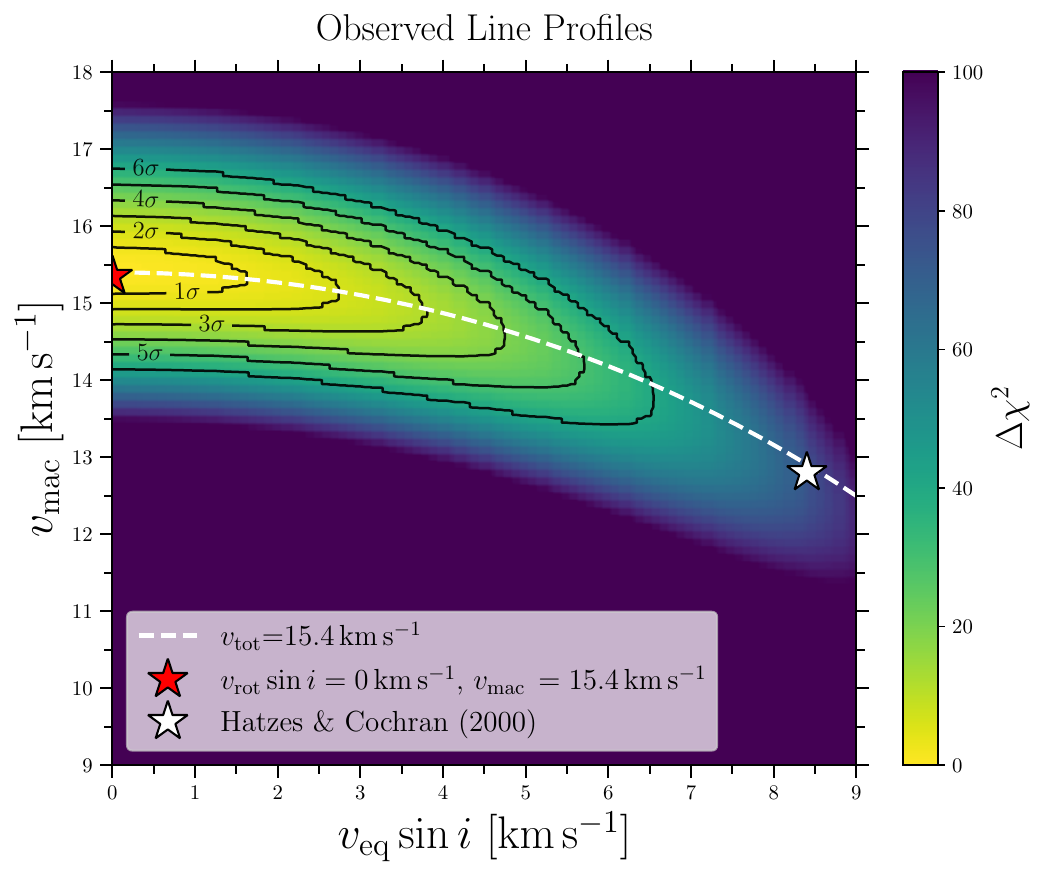}
    \end{minipage}\hfill
    \begin{minipage}{0.48\textwidth}
        \centering
        \includegraphics[width=\linewidth]{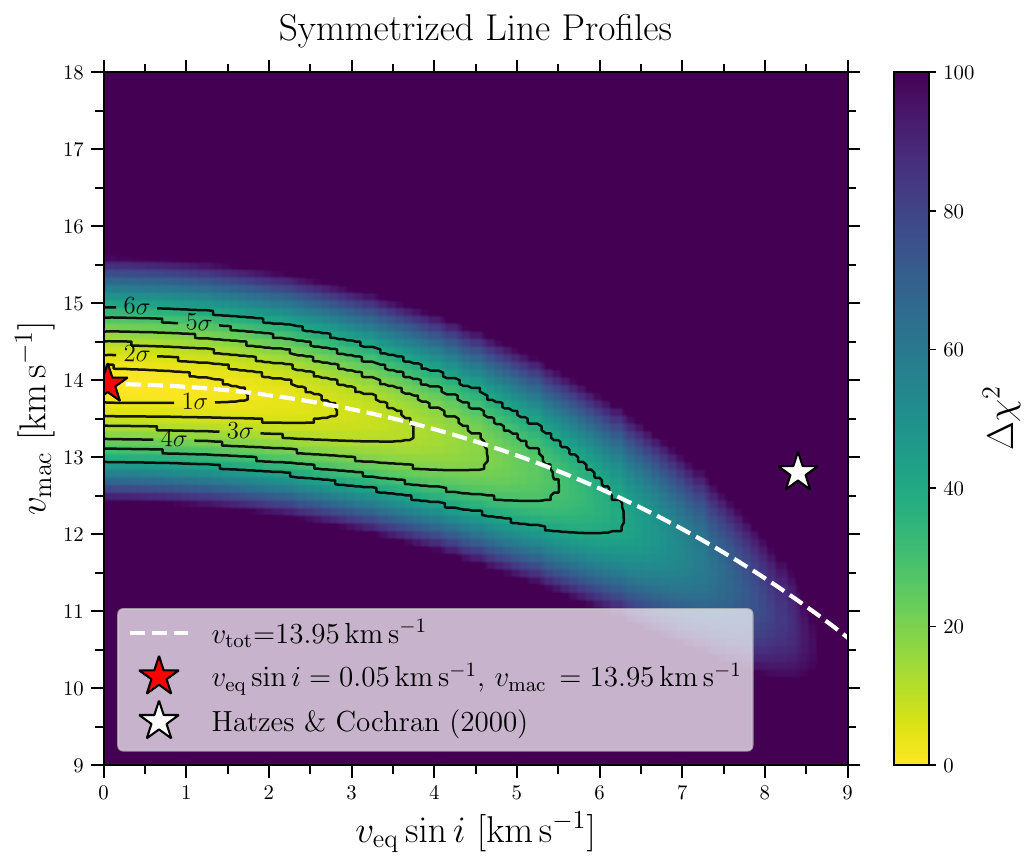}
    \end{minipage}
    \caption{$\Delta\chi^{2}$ as a function of $v_{\mathrm{eq}}\sin i_{\star}$ and $v_{\mathrm{mac}}$ from fits to the five observed (left) and symmetrized (right) metal line profiles.}
    \label{fig:chi2_obs}
\end{figure*}

To explore the correlation between $v_{\mathrm{eq}}\sin i_{\star}$ and $v_{\mathrm{mac}}$ we generated $\chi^{2}$ contour plots for the observed and symmetrized line profiles. For the $\chi^{2}$ calculations the $v_{\mathrm{mic}}$ and the elemental abundances were fixed to the values obtained from fitting the large spectral regions. Fixing these parameters allowed us to characterize the covariance between $v\sin i$ and $v_{\mathrm{mac}}$ without introducing additional degeneracies. The RV was again allowed to vary. Figure~\ref{fig:chi2_obs} shows the $\Delta\chi^{2}=(\chi^{2}-\chi^{2}_{\mathrm{min}})/\chi^{2}_{\nu,\mathrm{min}}$ surface for fits to the five observed and symmetrized line profiles. We normalized by $\chi^{2}_{\nu,\mathrm{min}}$ to help account for systematic differences between the model and observations. The best fit models to the five observed and five symmetrized line profiles minimize rotational broadening ($v_{\mathrm{eq}}\sin i_{\star}\approx0$\,km\,s$^{-1}$), giving $v_{\mathrm{mac}}\approx14-15$\,km\,s. However, a strong correlation between $v_{\mathrm{eq}}\sin i_{\star}$ and $v_{\mathrm{mac}}$ is observed in the $\Delta\chi^{2}$ plots, which roughly traces the total combined broadening $v_{\mathrm{tot}}=\sqrt{(v_{\mathrm{eq}}\sin i_{\star})^{2} + (v_{\mathrm{mac}})^{2}}$. 

To the best of our knowledge \cite{Hatzes_2000} provide the most recent literature measurement of Polaris' $v_{\mathrm{eq}}\sin i_{\star}$. The authors report $v_{\mathrm{eq}}\sin i_{\star}=8.4$\,km\,s$^{-1}$ and a macroturbulent velocity of $v_{\mathrm{mac}}=12.8$\,km\,s$^{-1}$ from a fit to the Sc\,\textsc{ii}\,$\lambda5526$ line from spectra with $R\sim50,000$. Our fitting of both the spectral regions and selected line profiles suggests low to absent rotational broadening, inconsistent with the findings of \cite{Hatzes_2000}. However, as shown in Fig~\ref{fig:chi2_obs} the broadening values given by \cite{Hatzes_2000} lie close to the same $v_{\mathrm{tot}}=15.4$\,km\,s$^{-1}$ curve as our best fit values. This suggests that the discrepancy between our results is due to modelling systematics from the assumption of the form for $v_{\mathrm{mac}}$. 

From our analysis we conclude that we cannot confidently disentangle the broadening parameters from $\chi^{2}$ fitting. To set upper limits on $v_{\mathrm{eq}}\sin i_{\star}$ we fixed $v_{\mathrm{mac}}=0$\,km\,s$^{-1}$ and fit the five observed and symmetrized line profiles obtaining $v_{\mathrm{eq}}\sin i_{\star}=13.5\pm0.1$\,km\,s$^{-1}$ and $v_{\mathrm{eq}}\sin i_{\star}=13.3\pm0.1$\,km\,s$^{-1}$ respectively. The model fits are shown in Fig.~\ref{fig:5lines_fit}. From these results we set a conservative upper bound of ${v_{\mathrm{eq}}\sin i_{\star}<13.5}$\,km\,s$^{-1}$.

From ${v_{\mathrm{eq}} = 23.3\pm0.2}$\,km\,s$^{-1}$ and ${v_{\mathrm{eq}}\sin i_{\star}<13.5}$\,km\,s$^{-1}$ we set a conservative upper bound of $i_{\star}<37^{\circ}$ on the stellar inclination angle (for $0^{\circ}\leq i_{\star}\leq90^{\circ}$). Assuming $v_{\mathrm{eq}}\sin i_{\star}=8.4$\,km\,s$^{-1}$ from \cite{Hatzes_2000} would imply a lower inclination of $\sim21^{\circ}$. 

\section{Binary Spin-Orbit Misalignment}\label{sec:spin_orbit}
We investigated our ability to set constraints on the spin orbit alignment of Polaris~Aa. We work in the standard coordinate system where $\hat{\boldsymbol{x}}$ and $\hat{\boldsymbol{y}}$ are in the tangent plane of the sky and $\hat{\boldsymbol{z}}$ points along the line of sight away from the observer. The unit normal vector to the orbital plane is given by
\begin{equation}\label{eq:orbit_normal}
    \hat{\boldsymbol{n}}_{\mathrm{o}}= \begin{bmatrix}
    \sin (i_{\mathrm{o}})\sin(\Omega_{\mathrm{o}}) \\
    -\sin(i_{\mathrm{o}})\cos(\Omega_{\mathrm{o}}) \\
    \cos(i_{\mathrm{o}})\end{bmatrix}
\end{equation}
where $i_{\mathrm{o}}$ is the orbital inclination angle and $\Omega_{\mathrm{o}}$ is the longitude of the ascending node \citep{Hilditch_2001}. The obliquity angle $\beta$ between ${\hat{\boldsymbol{n}}}_{\mathrm{o}}$ and the stellar rotation axis ${\hat{\boldsymbol{n}}}_{\star}$ is then given by
\begin{equation}\label{eq:beta}
\beta=\cos^{-1}\left({\hat{\boldsymbol{n}}}_{\mathrm{o}}\cdot\hat{\boldsymbol{n}}_{\star}\right).
\end{equation} 
We calculated the probability density function (PDF) of $\beta$ through Monte Carlo sampling. To sample ${\hat{\boldsymbol{n}}}_{\mathrm{o}}$ we adopted Gaussian priors for $i_{\mathrm{o}}$ and $\Omega_{\mathrm{o}}$ using the most likely values and 1$\sigma$ uncertainties from the ``replicated" weighted solution reported by \cite{Evans_2024}. We sampled ${\hat{\boldsymbol{n}}_{\star}=\begin{pmatrix} \sin\theta\cos\phi & \sin\theta\sin\phi & \cos\theta \end{pmatrix}}$ in spherical coordinates where $\theta$ is the polar angle and $\phi$ is the azimuthal angle. From our analysis in Sec.~\ref{sec:rot_and_inc} we have the constraint $i_{\star}<37^{\circ}$. Circular polarization alone cannot resolve the $180^{\circ}$ ambiguity on $i_{\star}$. As $i_{\mathrm{o}}\approx128^{\circ}$, ${\hat{\boldsymbol{n}}}_{\mathrm{o}}$ points towards the observer (Eq.~\ref{eq:orbit_normal}). Therefore $\hat{\boldsymbol{n}}_{\star}$ is prograde with respect to the orbit when $\theta\in[143^{\circ},\,180^{\circ}]$. We consider only the prograde solution in our analysis. A rotation axis retrograde to the orbit may be physically possible through extreme dynamical interactions such as orbit flipping by eccentric Kozai-Lidov cycles prior to a merger (see Section~\ref{sec:merger_hypothesis}). However, we lack detailed N-body modelling to put an informed prior on such a scenario.

\begin{figure}
    \centering
    \includegraphics[width=\linewidth]{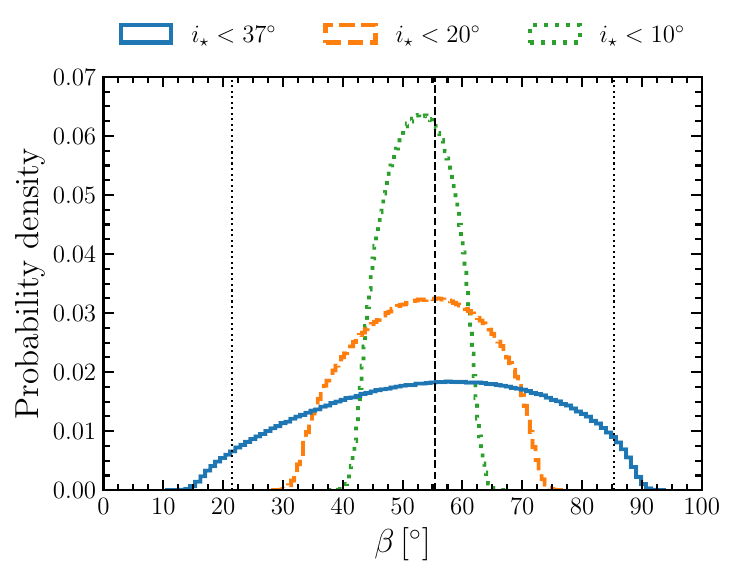}
    \caption{Probability density functions for the obliquity angle $\beta$ between the stellar rotation axis $\hat{\boldsymbol{n}}_{\star}$ and orbital normal axis ${\hat{\boldsymbol{n}}}_{\mathrm{o}}$ assuming the stellar rotation is prograde relative to the orbit. The solid blue line denotes the distribution for $i_{\star}<37^{\circ}$. The vertical dashed line denotes the median $\beta\approx55.4^{\circ}$ value of the distribution, and the vertical dotted lines denote the 95\% CI. The dashed orange and dotted green lines show PDFs assuming improved constraints on $i_{\star}$.}
    \label{fig:density_dist}
\end{figure}

We sampled $\cos\theta$ uniformly for $\theta$ over the range $[143^{\circ},\,180^{\circ}]$. For $\phi$ we sampled uniformly over the range $[0^{\circ},\,360^{\circ}]$. We generated $10^{6}$ samples of $\beta$ using Eq.~\ref{eq:beta} to estimate the PDF shown in Figure~\ref{fig:density_dist}. The PDF shows a high likelihood of a strong misalignment between $\hat{\boldsymbol{n}}_{\star}$ and ${\hat{\boldsymbol{n}}}_{\mathrm{o}}$ with median $\beta\approx55.4^{\circ}$ and a 95\% confidence interval (CI) of [$21.4^{\circ},\, 85.4^{\circ}$]. The PDF gives a lower bound of $\beta>18.7^{\circ}$ at 99\% confidence. 

 Our constraint on $i_{\star}$ is a conservative upper bound and may be improved in future work through magnetic mapping or interferometric imaging. In Figure~\ref{fig:density_dist} we also show PDFs assuming improved constraints of $i_{\star}<20^{\circ}$ and $i_{\star}<10^{\circ}$.

\section{Discussion} \label{sec:Discussion}
\subsection{Rotational Properties}\label{sec:rot_properties}
We have performed the first direct measurement of $P_{\mathrm{rot}}$ and $v_{\mathrm{eq}}$ for a Cepheid, demonstrating the power of magnetometry for determining the rotation properties of evolved, intermediate-mass and massive stars. Our measurement of $P_{\mathrm{rot}}$ is inconsistent with previously reported spectroscopic periods (e.g. \citealt{Barbey_2025}). This indicates that these signals have physical origins likely unrelated to rotation. While we do not detect any formally significant periods in the BIS measurements, a peak at $P_{\mathrm{rot}}$ is distinguishable in the GLS periodogram (Fig.~\ref{fig:GLS_BIS}). This suggests that the signal at $P_{\mathrm{rot}}$ may be present in long spectroscopic time series, but is weaker than other sources of variability. Although detailed evolutionary modelling is beyond the scope of this work, future studies should incorporate $v_{\mathrm{eq}}$ to help constrain the rotational history of Polaris. 

\subsection{Uncertain Origin of the Magnetic Field}\label{sec:uncertain_origin}
The magnetic properties of Polaris are challenging to reconcile with our current understanding of magnetism in intermediate-mass stars. The variability in Polaris' LSD~Stokes~$V$ profiles and $\langle B_{z}\rangle$ measurements has remained remarkably stable over five years of observations, spanning $\sim18$ rotation cycles and $\sim460$ pulsation cycles. The stability of Polaris' magnetic field is reminiscent of the so-called `fossil fields' observed in a subset ($<10\%$) of hot, main-sequence (MS) stars with spectral types O, B and A \citep{Morel_2015, Fossati_2015, Wade_2016, Sikora_2019a, Petit_2019}. Fossil magnetic fields tend to be strong (order 100\,G to 10\,kG), with predominantly dipolar geometries that are stable over decades \citep{Oksala_2012, Silvester_2014, Grunhut_2017, Shultz_2018, Sikora_2019b}. 

However, Polaris' Stokes~$V$ profiles do not resemble those of a simple dipole. The Stokes~$V$ profiles are complex with multiple polarity reversals across the $V/I_{C}=0$ axis. This morphology is qualitatively similar to the Stokes~$V$ signatures observed in low-mass MS stars, which commonly possess complex magnetic field geometries and convective dynamo driven variability cycles \citep{Folsom_2016, Folsom_2018, Donati_2023, Bellotti_2025}.

The apparent tension between the stability and complexity of Polaris' magnetic signature highlights our incomplete understanding of magnetism in evolved intermediate and massive stars. Specifically, the complex relationship between magnetic fields, convection, pulsation and changes in internal structure as stars transition across instability regions. 

As a $\sim5\,M_{\odot}$ F8Ib yellow-supergiant, Polaris' progenitor was presumably a mid B-type star on the MS, comprised of a convective core with a primarily radiative outer envelope and near surface He~{\sc ii} convection zone \citep{Cantiello_2019, Jermyn_2022}. As the star cooled and expanded, shallow H and He~{\sc i} convection zones formed, helping to drive the stellar pulsation in the IS \citep{Bono_1999, Stuck_2025}. \cite{Charbonnel_2017} do not predict an $\alpha-\omega$ dynamo at this evolutionary stage due to the inefficiency of the thin, outer convective envelope. After crossing the IS, an intermediate-mass star will develop a deep convective envelope on the red giant branch. Weak surface magnetic fields have been directly detected in red giants \citep{Auriere_2009, Auriere_2015, Borisova_2016, Tsvetkova_2017}, and are believed to be generated by convective dynamos \citep{Charbonnel_2017, Amard_2024}. 

After ignition of core helium burning, a $5\,M_{\odot}$ star will move back through the IS on a blue loop for second and third crossings, and the outer envelope will transition back from convective to primarily radiative. The impact of this transition on an envelope magnetic field is unclear. Persistence of magnetic fields between convective and radiative regimes is observed in other stellar contexts. Magnetic fields are detected in intermediate mass Herbig~Ae/Be and T~Tauri stars with a similar incidence rate to their MS counterparts \citep{Wade_2007, Alecian_2013, Villebrun_2019}. This suggests that magnetic fields generated during convective pre-main sequence stages may relax into stable fossil-field configurations on the MS \citep{Neiner_2015}. Strong magnetic fields ($>10^{5}$\,G) detected in the radiative cores of red giants through asteroseismic techniques are believed to be fossil-field remnants from dynamos that operated during the convective core stage on the MS \citep{Fuller_2015, Bugnet_2021}. There is also a small subset of red giants that are hypothesized to be the descendants of strongly magnetic Ap-type stars due to anomalously high activity levels and $\langle B_{z}\rangle$ strengths for their rotation rate \citep{Stepien_1993, Auriere_2015, Borisova_2016}. 

It is challenging to ascribe a firm origin for the magnetic field due to the ambiguity surrounding Polaris' crossing number. The predominantly positive rate of period change $\dot{P}$ over the last $\sim175$\,years \citep{Torres_2023} suggests an increasing $R_{\star}$ and a 1st or 3rd crossing. Authors have argued a 1st \citep{Turner_2013, Fadeyev_2015, Anderson_2018, Ripepi_2021} or 3rd  \citep{Neilson_2014, Evans_2018} crossing based on the luminosity, $\dot{P}$ and CNO abundances. Conflicting interpretations are partially due to the historical disagreement over the distance to Polaris~A \citep{Engle_2018}. If Polaris is on its 1st crossing than the magnetic field may be of fossil origin, or there is a dynamo mechanism efficient enough to generate and sustain the global field. Such a dynamo would be presumably generated in the sub-surface convective zones. If Polaris is on its 3rd crossing, then we should consider the possibility that a magnetic field generated during the red-giant phase can persist as a type of fossil field through blue loops. 

\subsection{Comparison with the Broader Cepheid Sample}
Polaris' magnetic signature is also challenging to interpret in the context of our larger magnetic survey of Cepheids. In addition to Polaris, \cite{Barron_2022} reported magnetic detections in $\delta$~Cep, $\eta$~Aql and $\zeta$~Gem. In contrast to Polaris, these targets show strongly asymmetric single- or double-lobed Stokes~$V$ profiles. We observe similar features in most other detected survey targets, which are the subject of a forthcoming paper (\citealt{Barron_2024}, in prep.). A notable exception is the complex Stokes~$V$ signature of MY~Pup \citep{Barron_2024}. Like Polaris, MY~Pup ($P_{\mathrm{puls}}=5.7$\,d; F7Ib/II) is a first overtone pulsator \citep{Kienzle_1999, Hocde_2024} with a pulsation amplitude larger than Polaris, but low relative to fundamental Cepheids with similar $P_{\mathrm{puls}}$. 

Stokes~$V$ asymmetries and shock enhancement have been observed in other evolved pulsators, including the Mira $\chi$~Cyg \citep{Lebre_2014, Lavail_2025} and the RV~Tauri stars R~Sct and U~Mon \citep{Sabin_2015, Georgiev_2023}. Notably, net positive circular polarization has also been detected in the rotationally averaged Stokes~$V$ profiles of three weakly magnetic Am stars \citep{Petit_2011, Blazere_2016}. As \cite{Blazere_2016} discuss, similarly anomalous Stokes~$V$ signatures are detected in solar magnetic regions and arise from velocity and magnetic field gradients along the line of sight \citep{lopez_2002, Martinez_2008, Bellot_2019}. In the case of Am stars \cite{Blazere_2016} suggest that the thin convective layer near the surface may generate the necessary velocity and magnetic field gradients \citep{Kupka_2009}. \cite{Folsom_2018_zeeman} demonstrated that inclusion of atmospheric gradients in polarized spectrum synthesis models can qualitatively reproduce Am star single lobed Stokes~$V$ signatures. Such a modelling approach has not yet been applied to Cepheids.

\subsection{Comparison with Non-Cepheid Supergiants}
Magnetic detections are rare among massive ($\gtrsim8\,M_{\odot}$) post-MS B and A-type supergiants \citep{Verdugo_2003, Aerts_2013, Shultz_2014}. The handful of magnetically detected A-type supergiants possess weak ($\sim1$\,G) $\langle B_{z}\rangle$ strengths and Stokes~$V$ signatures consistent with a dipole field, suggesting they may be the evolved descendants of fossil field hosting MS OB-type stars \citep{Neiner_2017, Martin_2018, Wade_2025}. At cooler temperatures, magnetic fields have been detected in a small number of non-Cepheid F, G and K-type supergiants \citep{Grunhut_2010}. Similar to the A supergiants, these stars have weak $\langle B_{z}\rangle$ strengths but present a wider variety of Stokes~$V$ morphologies. \cite{Wade_2025} recently presented magnetic monitoring of the F-type supergiants $\alpha$~Lep (F0Ib) and $\alpha$~Per (F5Ib). The authors suggested a dynamo origin for the fields due to the complexity and variability of the Stokes~$V$ signatures.

The Stokes~$V$ signatures of $\alpha$~Per bear strong resemblance to those of Polaris, both in their complexity and asymmetry about $V/I_{C}=0$. Previously suggested to be a low-amplitude Cepheid, $\alpha$~Per lies outside the blue edge of the IS \citep{Hatzes_1995} and presents multiple long periods of uncertain origin \citep{Lee_2012, Lee_2025, Wade_2025}. In light of our results, we suggest an alternative explanation to \cite{Wade_2025} for $\alpha$~Per's Stokes~$V$ variability. Like Polaris, $\alpha$~Per may host a complex but stable magnetic field configuration. In this scenario, the differences in Stokes~$V$ variability may be attributed to Polaris' low inclination angle relative to $\alpha$~Per (Sec.~\ref{sec:rot_and_inc}). More observations of $\alpha$~Per are needed to confidently determine a rotation period and test this hypothesis. Detections in additional targets are required to determine if the Stokes~$V$ morphology of $\alpha$~Per is typical of non-Cepheid mid to late F-type supergiants.

\subsection{Merger Hypothesis}\label{sec:merger_hypothesis}
\cite{Bond_2018} identified a large age discrepancy between the Cepheid Polaris~Aa and the wide companion Polaris B (F3\,V) from isochrone modelling in the $B-V$ colour magnitude diagram. \cite{Evans_2018}  estimate an approximate age of 100\,Myr for Polaris A and 2.25\,Gyr for Polaris B. Assuming that Polaris B is a physically associated, coeval companion of Polaris Aa, this tension may be explained if Polaris~Aa is a merger product \citep{Bond_2018, Evans_2018}.  

A merger scenario is interesting to consider in the context of Polaris' magnetic field. Stellar mergers appear to be a viable channel to produce some of the strongly magnetic O and B-type stars on the MS \citep{Schneider_2016, Schneider_2019, Vynatheya_2026}. As we discuss in Sec.~\ref{sec:uncertain_origin}, Polaris' Stokes~$V$ signatures do not resemble the simple dipole signatures commonly associated with fossil-field hosting progenitors. However, we have little observational evidence concerning the long-term stability of magnetic fields produced through binary interaction, and how fossil-fields respond to the development of envelope convection as the star evolves.  

A merger scenario may have also a significant impact on Polaris' rotational history. As \cite{Schneider_2025} discuss the immediate products of most stellar mergers can be assumed to be rotating near the break-up velocity. However, merged stars may spin down efficiently, losing angular momentum through their remnant disc and magnetic braking to become slow rotators. There is some observational evidence to support this, as \cite{Ferraro_2023} found that slower rotating blue stragglers prefer higher density environments with high collision rates. Merger products may also possess different interior structures than their single star counterparts, complicating evolution modelling \citep{Bellinger_2024, Henneco_2024}. 
 
The spin-orbit misalignment of Polaris~Aa may be consistent with a past dynamical interaction. As \cite{Anderson_2018} discuss, the high eccentricity of the Aa+Ab binary suggests past Lidov-Kozai interactions. In a merger scenario, the Polaris system would have begun as a hierarchical quadruple with an inner binary. Kozai-Lidov oscillations could have driven a tightening of the inner binary orbit, leading to a stellar merger \citep{Perets_2009, Naoz_2014}. For a hierarchical system subject to the eccentric Kozai-Lidov effect \citep{Naoz_2016}, the inner binary orbit may even flip orientation with respect to the outer perturber \citep{Naoz_2013}. Further investigation of Polaris as a potential merger product should include detailed evolutionary modelling and N-body simulation. 

\subsection{Future Work}
Future work should explore reconstruction of the large-scale magnetic field geometry through inversion methods such as Zeeman Doppler Imaging (ZDI; \citealt{Kochukhov_2016}). A magnetic mapping may aid in further constraining the stellar inclination angle and in turn the degree of spin-orbit misalignment. Additionally, a magnetic map could be compared to interferometric surface reconstructions \citep{Evans_2024} to search for correlations between the distribution of surface spots and the magnetic field geometry. However, reconstructing the surface field presents several challenges. Additional spectropolarimetric observations are needed over some rotational phases, particularly between $\phi_{\mathrm{rot}}=0$ and $\phi_{\mathrm{rot}}=0.3$ (Fig.~\ref{fig:LSD_StokesIVN}). To perform a magnetic inversion, the stellar pulsation and asymmetry of the LSD Stokes~$V$ profiles will likely need to be considered. 

Long-term magnetic monitoring of Polaris is necessary to examine the field's stability and obtain further insight into the field's origins. Continued observation will also probe the possible relation between magnetic field strength and changing pulsation amplitude due to magnetoconvective cycles \citep{Stothers_2009}.

A key challenge in our study of Polaris has been constraints on the frequency and duration of ESPaDOnS mountings due to competitive pressure from other CFHT instruments. These limitations will be partially alleviated with the upcoming installation of the Wenaokeao optomechanical interface. Wenaokeao will provide simultaneous mounting of ESPaDOnS and the SPIRou near-infrared spectropolarimeter, increasing the number of nights ESPaDOnS is mounted per semester. We expect that this will aid in the continued study of Polaris and other slowly rotating stars including blue, yellow and red supergiants.

\section{Conclusions}\label{sec:Conclusions}
We analyzed 42 optical spectropolarimetric observations of Polaris obtained with ESPaDOnS between 2020 and 2025. Here we list the key conclusions of our study.  
\begin{enumerate}
    \item A complex LSD~Stokes~$V$ magnetic signature is detected in all spectropolarimetric observations. The Stokes~$V$ profiles are not strongly variable over single pulsation cycles, but are variable over long time spans.
    \item The $\langle B_{z}\rangle$ measurements vary between $-3$\,G and $+0.6$\,G. The $\langle B_{z}\rangle$ measurements are obtained at high precision with typical uncertainties of 
    ${\sim0.3}$\,G.
    \item From a period analysis of the $\langle B_{z}\rangle$ measurements we infer a rotation period of $P_{\mathrm{rot}}=100.29\pm0.19$\,days. No other significant periods are detected in the $\langle B_{z}\rangle$ measurements. The $\langle B_{z}\rangle$ measurements vary sinusoidally when phased to $P_{\mathrm{rot}}$, and the variability is stable over the 5\,year observing span. 
    \item From $P_{\mathrm{rot}}$ and previous determination of $\langle R_{\star}\rangle$ we obtain a precise measurement of the equatorial rotation velocity $v_{\mathrm{eq}}=23.3\pm0.2$\,km\,s$^{-1}$.
    \item We are unable to confidently disentangle spectral line broadening contributions from $v_{\mathrm{eq}}\sin i_{\star}$ and $v_{\mathrm{mac}}$. We set a conservative upper limit of ${v_{\mathrm{eq}}\sin i_{\star}<13.5}$\,km\,s$^{-1}$. This constrains the stellar inclination to $i_{\star}<37^{\circ}$.
    \item We set a lower bound of $\beta>18.7^{\circ}$ on the obliquity between the stellar rotation and orbital axes. There is a high probability that the axes are strongly misaligned. 
    \item The origin of Polaris' magnetic field remains unclear. The apparent tension between the complexity of the Stokes~$V$ signatures and long term stability of the $\langle B_{z}\rangle$ measurements makes it challenging to attribute Polaris' magnetic field to a fossil-field or $\alpha-\omega$ dynamo model. 
    \item Future work should explore reconstruction of the surface magnetic field geometry. A magnetic map could be compared to interferometric surface reconstructions to look for correlations between the magnetic field geometry and surface features.
\end{enumerate}

\begin{acknowledgments}
We thank the anonymous referee for helpful comments which led to the improvement of this manuscript. We are grateful to Dr. Nadine Manset for assistance with scheduling the observations and questions regarding data reduction. We thank Dr. Richard Anderson for helpful discussions.  

J.A.B. acknowledges support through a Postgraduate Doctoral Scholarship (PGS D) from the Natural Sciences and Engineering Research Council (NSERC) of Canada. G.A.W. acknowledges support in the form of a Discovery Grant from the Natural Sciences and Engineering Research Council (NSERC) of Canada. C.P.F. acknowledges funding from the European Union’s Horizon Europe research and innovation programme under grant agreement No. 101079231 (EXOHOST), and from the United Kingdom Research and Innovation (UKRI) Horizon Europe Guarantee Scheme (grant number 10051045).

Based on observations obtained at the Canada-France-Hawai`i 
Telescope (CFHT) which is operated by the National Research Council of  Canada, the Institut National des Sciences de l'Univers of the Centre  National de la Recherche Scientifique of France, and the University of  Hawai`i. CFHT is located on Maunakea on Hawai`i Island, a mountain of considerable cultural, natural, and ecological significance. Maunakea is a sacred site to Native Hawaiians, also known as K\=anaka `\=Oiwi. We would like to thank the Canada-France-Hawai`i Telescope (CFHT)  Operations and Software Groups for their contributions and diligence in  maintaining observatory operations; the CFHT Astronomy Group for their  observation coordination and data acquisition efforts; and the CFHT Finance \& Administration Group for their contributions to the management and administration of the observatory.

This research used the facilities of the Canadian Astronomy Data Centre operated by the National Research Council of Canada with the support of the Canadian Space Agency. This work has made use of the VALD database, operated at Uppsala University, the Institute of Astronomy RAS in Moscow, and the University of Vienna. This research has made use of the SIMBAD database, operated at CDS, Strasbourg, France, and NASA’s Astrophysics Data System (ADS).
\end{acknowledgments}





%
\facilities{CFHT (ESPaDOnS)}

\software{
Astropy \citep{astropy_2022},
corner \citep{corner},
emcee \citep{emcee_2013},
Matplotlib \citep{Hunter_2007}, 
NumPy \citep{Harris_2020},
SciPy \citep{SciPy_2020},
SpecpolFlow \citep{Folsom_2025},
ZEEMAN \citep{Landstreet_1988, Wade_2001}}


\appendix
\FloatBarrier
\setcounter{table}{0}
\setcounter{figure}{0}

\renewcommand{\thetable}{\Alph{section}\arabic{table}}
\renewcommand{\thefigure}{\Alph{section}\arabic{figure}}

\renewcommand{\theHtable}{\Alph{section}\arabic{table}}
\renewcommand{\theHfigure}{\Alph{section}\arabic{figure}}

\FloatBarrier
\section{Observation Log}
Table~\ref{tab:obs} provides a log of the ESPaDOnS observations of Polaris, including corresponding $\langle B_{z}\rangle$ and RV measurements, as well as pulsation and rotation phase values. 
\begin{deluxetable}{rrrrrrrrrr}
 \tabletypesize{\footnotesize}
 \caption{Log of ESPaDOnS spectropolarimetric observations of Polaris.}
\startdata
\tablehead{\colhead{Date} & \colhead{HJD\,$-$}  & \colhead{Exp. Time}  & \colhead{S/N}  & \colhead{$\langle B_{z}\rangle$}  & \colhead{RV$_{\mathrm{cog}}$} & \colhead{RV$_{\mathrm{corr}}$} & \colhead{$\phi_{\textrm{puls}}$} & \colhead{$\phi_{\textrm{rot}}$} & \colhead{$N_{\mathrm{rot}}$} \\
\colhead{} & \colhead{2\,400\,000} & \colhead{(s)} & \colhead{(500\,nm)} & \colhead{(G)} & \colhead{(km\,s$^{-1}$)} & \colhead{(km\,s$^{-1}$)} & \colhead{} & \colhead{} & \colhead{}}
2020-11-21 & 59174.7863 & $56\times4\times11$ & 6640 & $0.27\pm0.13$ & $-16.23\pm0.04$ & $-1.47\pm0.06$ & $0.143\pm0.003$ & $0.98\pm0.02$ & 1 \\
2021-08-27 & 59453.9480 & $12\times4\times11$ & 3190 & $-1.08\pm0.28$ & $-15.92\pm0.04$ & $-0.72\pm0.06$ & $0.426\pm0.002$ & $0.76\pm0.01$ & 4 \\
2021-08-29 & 59456.0836 & $12\times4\times11$ & 2780 & $-0.69\pm0.31$ & $-14.84\pm0.03$ & $0.37\pm0.06$ & $0.964\pm0.002$ & $0.78\pm0.01$ & 4 \\
2021-08-31 & 59457.9845 & $11\times4\times11$ & 2820 & $-0.32\pm0.32$ & $-15.73\pm0.04$ & $-0.52\pm0.06$ & $0.442\pm0.002$ & $0.80\pm0.01$ & 4 \\
2021-09-02 & 59460.1018 & $12\times4\times11$ & 3330 & $-0.56\pm0.26$ & $-14.99\pm0.03$ & $0.23\pm0.06$ & $0.975\pm0.002$ & $0.82\pm0.01$ & 4 \\
2021-11-23 & 59541.8165 & $8\times4\times11$ & 2360 & $-2.24\pm0.36$ & $-14.96\pm0.04$ & $0.38\pm0.06$ & $0.548\pm0.002$ & $0.64\pm0.01$ & 5 \\
2021-11-24 & 59542.8842 & $11\times4\times11$ & 2510 & $-2.28\pm0.36$ & $-13.88\pm0.03$ & $1.46\pm0.06$ & $0.817\pm0.002$ & $0.65\pm0.01$ & 5 \\
2021-11-26 & 59544.8267 & $12\times4\times11$ & 3050 & $-1.48\pm0.29$ & $-17.06\pm0.04$ & $-1.72\pm0.06$ & $0.306\pm0.002$ & $0.67\pm0.01$ & 5 \\
2022-02-17 & 59627.7469 & $12\times4\times11$ & 2790 & $-2.41\pm0.31$ & $-17.02\pm0.04$ & $-1.57\pm0.05$ & $0.182\pm0.002$ & $0.50\pm0.01$ & 6 \\
2022-02-19 & 59629.7593 & $12\times4\times11$ & 2890 & $-2.16\pm0.30$ & $-13.78\pm0.03$ & $1.67\pm0.05$ & $0.689\pm0.002$ & $0.52\pm0.01$ & 6 \\
2022-02-21 & 59631.7122 & $12\times4\times11$ & 3020 & $-3.08\pm0.29$ & $-16.96\pm0.04$ & $-1.51\pm0.05$ & $0.181\pm0.002$ & $0.53\pm0.01$ & 6 \\
2022-02-22 & 59632.7124 & $12\times4\times11$ & 3700 & $-2.50\pm0.25$ & $-16.11\pm0.04$ & $-0.65\pm0.05$ & $0.433\pm0.002$ & $0.54\pm0.01$ & 6 \\
2022-02-23 & 59633.7124 & $12\times4\times11$ & 3210 & $-3.04\pm0.28$ & $-13.87\pm0.03$ & $1.59\pm0.05$ & $0.684\pm0.002$ & $0.55\pm0.01$ & 6 \\
2022-07-07 & 59768.1199 & $11\times4\times11$ & 3180 & $-0.04\pm0.27$ & $-15.46\pm0.04$ & $0.18\pm0.06$ & $0.523\pm0.002$ & $0.89\pm0.01$ & 7 \\
2022-07-08 & 59769.1235 & $12\times4\times11$ & 3520 & $0.13\pm0.26$ & $-13.84\pm0.04$ & $1.80\pm0.05$ & $0.776\pm0.002$ & $0.90\pm0.01$ & 7 \\
2022-07-10 & 59771.0531 & $12\times4\times11$ & 3730 & $0.19\pm0.26$ & $-17.46\pm0.04$ & $-1.82\pm0.06$ & $0.262\pm0.002$ & $0.92\pm0.01$ & 7 \\
2022-07-11 & 59772.0497 & $12\times4\times11$ & 2330 & $-0.16\pm0.38$ & $-15.58\pm0.04$ & $0.06\pm0.05$ & $0.513\pm0.002$ & $0.93\pm0.01$ & 7 \\
2022-07-12 & 59773.1041 & $12\times4\times11$ & 3890 & $-0.03\pm0.23$ & $-13.82\pm0.03$ & $1.82\pm0.05$ & $0.778\pm0.002$ & $0.94\pm0.01$ & 7 \\
2022-07-14 & 59775.0434 & $12\times4\times11$ & 3690 & $0.33\pm0.25$ & $-17.42\pm0.04$ & $-1.78\pm0.05$ & $0.266\pm0.001$ & $0.96\pm0.01$ & 7 \\
2022-07-15 & 59776.0796 & $12\times4\times11$ & 3210 & $0.34\pm0.27$ & $-15.37\pm0.04$ & $0.28\pm0.06$ & $0.527\pm0.001$ & $0.97\pm0.01$ & 7 \\
2022-08-13 & 59805.0415 & $11\times4\times11$ & 1250 & $-1.34\pm0.72$ & $-13.90\pm0.04$ & $1.78\pm0.05$ & $0.819\pm0.001$ & $0.26\pm0.01$ & 8 \\
2022-08-14 & 59805.9626 & $11\times4\times11$ & 2890 & $-1.24\pm0.29$ & $-16.28\pm0.04$ & $-0.60\pm0.05$ & $0.051\pm0.001$ & $0.27\pm0.01$ & 8 \\
2022-09-09 & 59832.1341 & $11\times4\times11$ & 2780 & $-2.86\pm0.33$ & $-14.39\pm0.04$ & $1.33\pm0.05$ & $0.640\pm0.001$ & $0.53\pm0.01$ & 8 \\
2022-09-12 & 59834.9742 & $11\times4\times11$ & 2880 & $-2.62\pm0.31$ & $-17.10\pm0.05$ & $-1.38\pm0.06$ & $0.355\pm0.001$ & $0.56\pm0.01$ & 8 \\
2022-09-14 & 59837.0139 & $11\times4\times11$ & 3220 & $-2.14\pm0.28$ & $-14.33\pm0.04$ & $1.39\pm0.05$ & $0.868\pm0.001$ & $0.58\pm0.01$ & 8 \\
2022-09-15 & 59837.9894 & $11\times4\times11$ & 2550 & $-2.27\pm0.34$ & $-16.95\pm0.04$ & $-1.23\pm0.05$ & $0.114\pm0.001$ & $0.59\pm0.01$ & 8 \\
2022-10-07 & 59860.0585 & $11\times4\times11$ & 3660 & $-0.71\pm0.25$ & $-14.12\pm0.04$ & $1.63\pm0.05$ & $0.670\pm0.001$ & $0.81\pm0.01$ & 8 \\
2022-10-15 & 59868.1379 & $11\times4\times11$ & 3890 & $0.27\pm0.25$ & $-14.04\pm0.03$ & $1.72\pm0.05$ & $0.704\pm0.001$ & $0.89\pm0.01$ & 8 \\
2022-10-19 & 59871.9739 & $11\times4\times11$ & 3520 & $0.20\pm0.25$ & $-14.25\pm0.04$ & $1.51\pm0.05$ & $0.670\pm0.001$ & $0.93\pm0.01$ & 8 \\
2023-06-24 & 60120.0910 & $11\times4\times11$ & 3280 & $-2.27\pm0.29$ & $-17.45\pm0.04$ & $-1.41\pm0.05$ & $0.137\pm0.001$ & $0.40\pm0.01$ & 11 \\
2023-07-08 & 60134.0471 & $11\times4\times11$ & 3540 & $-2.56\pm0.25$ & $-14.53\pm0.03$ & $1.53\pm0.05$ & $0.651\pm0.001$ & $0.54\pm0.01$ & 11 \\
2023-10-20 & 60237.9519 & $10\times4\times11$ & 2500 & $-2.87\pm0.36$ & $-14.36\pm0.03$ & $1.80\pm0.05$ & $0.810\pm0.001$ & $0.58\pm0.01$ & 12 \\
2023-10-26 & 60243.8551 & $11\times4\times11$ & 2700 & $-2.09\pm0.34$ & $-17.91\pm0.04$ & $-1.74\pm0.05$ & $0.296\pm0.001$ & $0.64\pm0.01$ & 12 \\
2023-12-04 & 60282.7335 & $11\times4\times11$ & 3000 & $0.67\pm0.29$ & $-17.25\pm0.04$ & $-1.04\pm0.05$ & $0.085\pm0.001$ & $0.03\pm0.01$ & 13 \\
2024-01-01 & 60310.8111 & $11\times4\times11$ & 3650 & $-1.69\pm0.27$ & $-17.85\pm0.04$ & $-1.61\pm0.05$ & $0.154\pm0.001$ & $0.31\pm0.01$ & 13 \\
2024-01-06 & 60315.7430 & $10\times4\times11$ & 2390 & $-2.27\pm0.38$ & $-17.41\pm0.04$ & $-1.17\pm0.05$ & $0.395\pm0.001$ & $0.36\pm0.01$ & 13 \\
2025-01-16 & 60691.9370 & $10\times4\times11$ & 2920 & $0.33\pm0.31$ & $-17.92\pm0.04$ & $-1.33\pm0.05$ & $0.107\pm0.002$ & $0.11\pm0.01$ & 17 \\
2025-08-09 & 60897.0520 & $10\times4\times11$ & 2160 & $-0.10\pm0.40$ & $-14.52\pm0.04$ & $2.24\pm0.05$ & $0.748\pm0.002$ & $0.15\pm0.02$ & 19 \\
2025-08-11 & 60899.0570 & $11\times4\times11$ & 1900 & $0.24\pm0.46$ & $-18.61\pm0.04$ & $-1.85\pm0.05$ & $0.253\pm0.002$ & $0.17\pm0.02$ & 19 \\
2025-12-06 & 61015.9453 & $10\times4\times11$ & 2660 & $-2.23\pm0.32$ & $-15.02\pm0.03$ & $1.83\pm0.04$ & $0.681\pm0.003$ & $0.34\pm0.02$ & 20 \\
2025-12-09 & 61018.8354 & $11\times4\times11$ & 2980 & $-2.33\pm0.29$ & $-18.09\pm0.04$ & $-1.24\pm0.05$ & $0.409\pm0.003$ & $0.37\pm0.02$ & 20 \\
2025-12-10 & 61019.9171 & $11\times4\times11$ & 2830 & $-2.45\pm0.31$ & $-15.05\pm0.03$ & $1.80\pm0.04$ & $0.681\pm0.003$ & $0.38\pm0.02$ & 20 \\
\label{tab:obs}
\enddata
\end{deluxetable}

\FloatBarrier
\section{MCMC Analysis}\label{sec:MCMC}
\FloatBarrier
\begin{deluxetable}{rrr}
 \tabletypesize{\footnotesize}
 \caption{Model priors and posterior estimates for fit to $\mathrm{RV}_{\mathrm{cog}} - \mathrm{RV}_{\mathrm{orb}}$.}
\tablehead{\colhead{Parameter} & \colhead{Prior}  & \colhead{Value}} 
\startdata
$C$ [km\,s$^{-1}$]& $\mathcal{U}(-2,\,0)$ &$-1.247^{+0.008}_{-0.008}$\\
$A$ [km\,s$^{-1}$]& $\mathcal{U}(-3,\,0)$ & $-1.87^{+0.01}_{-0.01}$\\
$P_{\mathrm{puls}}$ [d]& $\mathcal{U}(3.969,\,3.975)$ & $3.97197^{+0.00004}_{-0.00004}$\\
$t_{\mathrm{0}}$ [$\mathrm{HJD}-2460000$]& $\mathcal{U}(95.50,\,95.90)$ & $95.715^{+0.005}_{-0.005}$
\label{tab:mcmc_RV}
\enddata
\end{deluxetable}

\begin{deluxetable}{rrr}
 \tabletypesize{\footnotesize}
 \caption{Model priors and posterior estimates for fit to $\langle B_{z}\rangle$.}
\tablehead{\colhead{Parameter} & \colhead{Prior}  & \colhead{Value}} 
\startdata
$B_{0}$ [G]& $\mathcal{U}(-3,\,1)$ &$-1.15^{+0.06}_{-0.06}$\\
$B_{\mathrm{1}}$ [G]& $\mathcal{U}(0,\,4)$ & $1.52^{+0.06}_{-0.05}$\\
$P_{\mathrm{rot}}$ [d]& $\mathcal{U}(95,\,105)$ & $100.29^{+0.19}_{-0.19}$\\
$t_{\mathrm{0}}$ [$\mathrm{HJD}-2460000$]& $\mathcal{U}(t_{\mathrm{mid}}\pm\frac{1}{2}P_{\mathrm{rot}})$ & $79.59^{+0.83}_{-0.83}$
\label{tab:mcmc_Bz}
\enddata
\end{deluxetable}

We performed an MCMC analysis using the \texttt{emcee} Python package \citep{emcee_2013} to estimate the parameters and associated uncertainties for Eq.~\ref{eq:rv_model} fit to the $\mathrm{RV}_{\mathrm{cog}} - \mathrm{RV}_{\mathrm{orb}}$ values (Sec.~\ref{sec:Pulsation_Period_Amplitude}) and Eq.~\ref{eq:cosine} fit to the $\langle B_{z}\rangle$ measurements (Sec.~\ref{sec:rot_period}). We adopt uniform priors for all parameters, which are provided in Tables~\ref{tab:mcmc_RV} and \ref{tab:mcmc_Bz}. For both Eqs.~\ref{eq:rv_model} and \ref{eq:cosine}, the prior for the epoch $t_{0}$ was constrained to be near the midpoint of the observations ($t_{\mathrm{mid}} = 2460097.3517$\,HJD). We performed the sampling with the default `stretch move' ensemble method \citep{Goodman_2010} using 100 walkers and $10^{5}$ steps. We verified convergence by ensuring that the chain length was greater than 50 times the autocorrelation time. The first 1000 steps were then discarded, and the chains were thinned by a factor of 100 to reduce correlations.

\begin{figure}
    \centering
    \includegraphics[width=\linewidth]{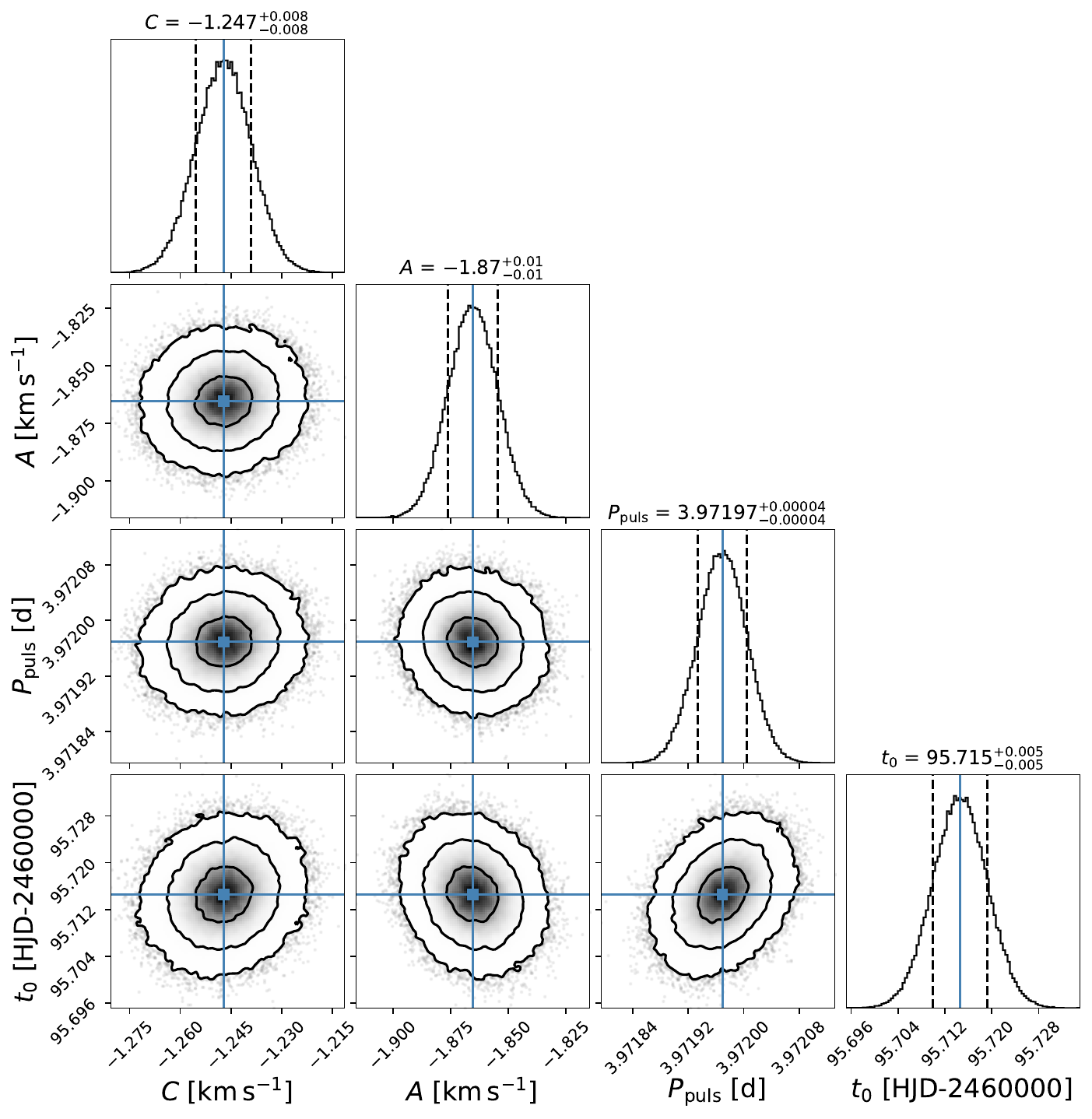}
    \caption{Marginalized posterior distributions inferred from fitting the $\mathrm{RV}_{\mathrm{cog}}-\mathrm{RV}_{\mathrm{orb}}$ measurements. The solid blue lines denote the median values and the dashed lines correspond to the 1$\sigma$ credible range. The 2D contours correspond to the $1\sigma$, $2\sigma$ and 3$\sigma$ joint credible regions.}
    \label{fig:corner_plot_RV}
\end{figure}

\begin{figure}
    \centering
    \includegraphics[width=\linewidth]{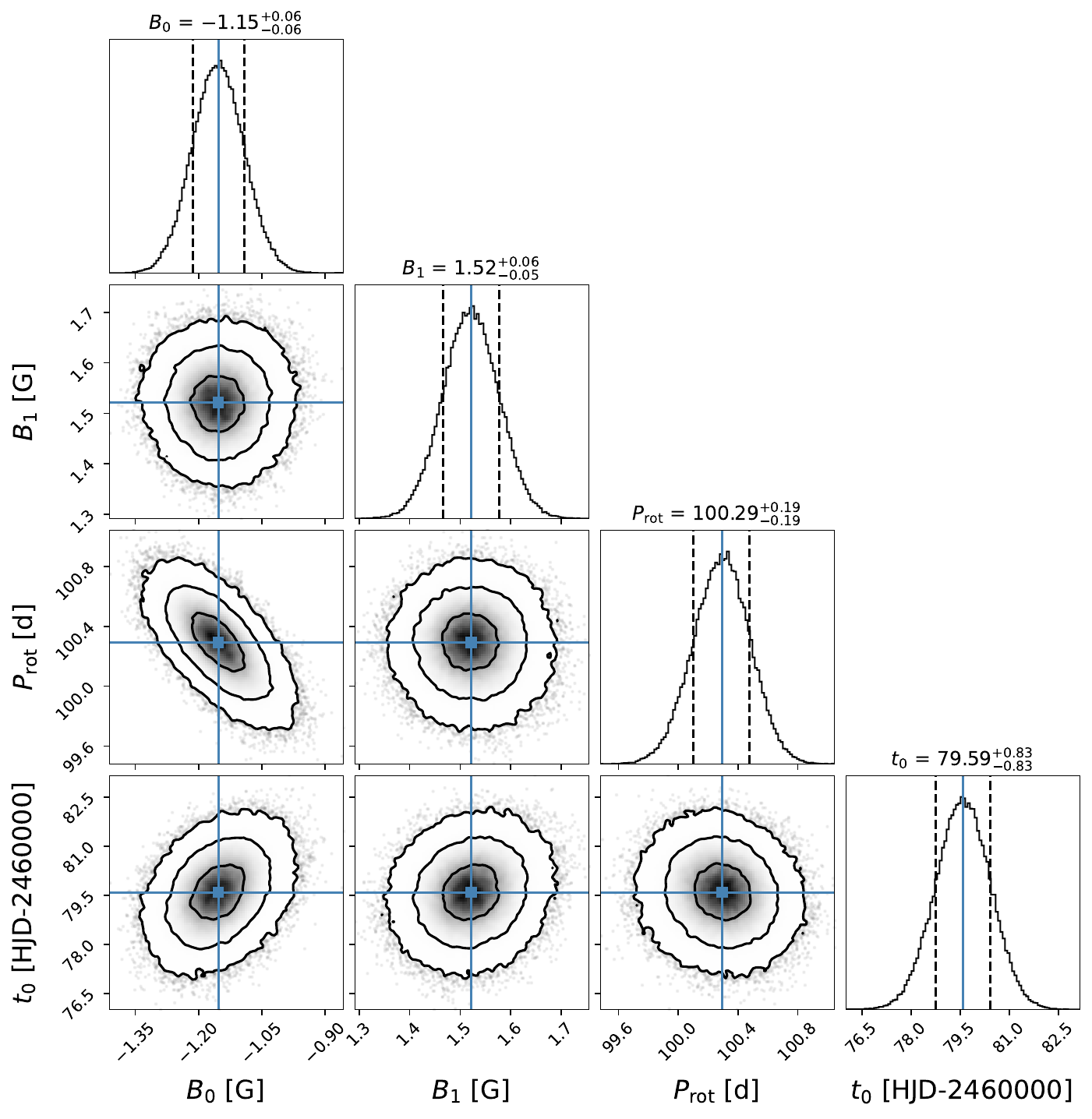}
    \caption{Marginalized posterior distributions inferred from fitting the $\langle B_{z} \rangle$ measurements. The solid blue lines denote the median values and the dashed lines correspond to the 1$\sigma$ credible range. The 2D contours correspond to the $1\sigma$, $2\sigma$ and 3$\sigma$ joint credible regions.}
    \label{fig:corner_plot_Bz}
\end{figure}

Figures~\ref{fig:corner_plot_RV} and \ref{fig:corner_plot_Bz}, and show the marginalized posterior distributions plotted with the \texttt{corner} Python package \citep{corner}. The parameter values reported in Tables~\ref{tab:mcmc_RV} and \ref{tab:mcmc_Bz} are the median of the marginalized posterior distributions, and the associated 1$\sigma$ uncertainties are the 16th and 84th percentiles.

\FloatBarrier
\bibliography{sample701}{}
\bibliographystyle{aasjournalv7}



\end{document}